\documentclass[preprint,12pt]{elsarticle}
\usepackage{amssymb}
\usepackage{rotating}

\usepackage{graphics}
\graphicspath{{figs/}}

\def\Journal#1#2#3#4{{#1} {\bf #2}, #3 (#4).}
\def\NIM{Nucl. Instr. and Meth.}

\def\NIMR{{Nucl. Instr. and Meth. in Phys. Res.} A}

\def\IEEE{IEEE Trans. Nucl. Sc.}
\def\be{\begin{equation}}
\def\ee{\end{equation}}
\newcommand{\ud}{\mathrm{d}}

\journal{Nuclear Instruments and Methods A}

\begin{document}
%\linenumbers

\begin{frontmatter}

\title{Transition Radiation Detectors}

\author[gsi]{A. Andronic}
\author[mun]{J.P. Wessels}

\address[gsi]{GSI Helmholtzzentrum f\"ur Schwerionenforschung,
D-64291 Darmstadt, Germany}
\address[mun]{Institut f\"ur Kernphysik, Universit\"at M\"unster, D-48149
  M\"unster, Germany; European Organization for Nuclear Research CERN, 1211
  Geneva, Switzerland}

\begin{abstract}
We review the basic features of transition radiation and how they are
used for the design of modern Transition Radiation Detectors
(TRD). The discussion will include the various realizations of
radiators as well as a discussion of the detection media and aspects
of detector construction. With regard
to particle identification we assess the different methods for
efficient discrimination of different particles and outline the
methods for the quantification of this property. Since a number of
comprehensive reviews already exist, we predominantly focus on
the detectors currently operated at the LHC. To a lesser extent we
also cover some other TRDs, which are planned or are currently being
operated in balloon or space-borne astro-particle physics experiments.
\end{abstract}

\begin{keyword}
\end{keyword}

\end{frontmatter}

%\vspace{.5cm}
%\date{\today}

\section{Introduction}\label{sect:intro}

%%%%%%%%%%%%%%%%%%%%%%%%%%%%%%%%%%%%%%%%%%%
In general, the interaction of a charged particle with a medium can
be derived from the treatment of its electromagnetic interaction with
that medium, where the interaction is mediated by a corresponding
photon. The processes that occur are ionization, Bremsstrahlung, Cherenkov 
radiation, and, in case of inhomogeneous media, transition radiation (TR). 
The latter process had been predicted by Ginzburg and Frank \cite{gin} in
1946. It was first observed in the optical domain by Goldsmith and
Jelley \cite{gold} in 1959 and further studied experimentally with
electron beams of tens of keV \cite{tr_opt}.
The relevance of this phenomenon for
particle identification went unnoted until it was realized that, for 
highly-relativistic charged particles ($\gamma\gtrsim 1000$), the spectrum 
of the emitted radiation extends into the X-ray domain \cite{gar}. 
While the emission probability for such an X-ray photon is small, its 
conversion leads to a large energy deposit compared to the average energy 
deposit via ionization. This led to the application of TR 
for particle identification at high momenta \cite{che0}.

Since then many studies have been pursued, both at the level of
the basic understanding of TR production \cite{ter,che1,art,dur} 
as well as with regard to the applications in particle detection and
identification \cite{che1,fab,cam,pri,cob,fab1,che2,fab2,bun}.
Consequently, TRDs have been used and are currently being used or
planned in a wide range of accelerator-based experiments, such as 
UA2 \cite{ans}, ZEUS \cite{zeus}, NA31 \cite{na31}, PHENIX \cite{ed1,ed2}, 
HELIOS \cite{dol}, D$\emptyset$ \cite{det,d0}, kTeV \cite{ktev}, 
H1 \cite{gra,h1}, WA89 \cite{wa89}, NOMAD \cite{nom1}, HERMES \cite{her}, 
HERA-B \cite{herab}, ATLAS \cite{atl}, ALICE \cite{ali}, CBM \cite {cbm}
and in astro-particle and cosmic-ray experiments:
WIZARD \cite{wizard}, HEAT \cite{heat}, MACRO \cite{macro},
AMS \cite {ams}, PAMELA \cite {pam}, ACCESS \cite{access}.
In these experiments the main purpose of the TRD is the discrimination 
of electrons from hadrons, but pion identification has been performed 
at Fermilab in a 250 GeV hadron beam \cite{errede} and $\pi/\Sigma$ 
identification has been achieved in a hyperon beam at CERN \cite{wa89}.

The subject of transition radiation and how it can be applied to
particle identification has already been comprehensively reviewed in
Ref. \cite{dol,fav}. An excellent concise review is given in \cite{pdg}.
Therefore, we restrict ourselves to a general description of the phenomenon 
and how TRD is employed in particle identification detectors. 
We will then concentrate on more recent developments of TRDs and specific 
analysis techniques, in particular for the detectors at the CERN 
Large Hadron Collider (LHC).

\section{Production of Transition Radiation}\label{sect:tr_prod}
%%%%%%%%%%%%%%%%%%%%%%%%%%%%%%%%%%%%%%%%%%%

\subsection{TR production in single foil radiators}

The practical theory of TR production is extensively presented in
References~\cite{che1,art,dur}. Extensions of the theory for
non-relativistic particles are covered in~\cite{hiro}. 
Here, we briefly summarize the most
important results for relativistic charged particles.

The double differential energy spectrum radiated by a charged particle with 
a Lorentz factor $\gamma$ traversing an interface between two dielectric 
media (with dielectric constants $\epsilon_1$ and $\epsilon_2$) 
has the following expression:
\be \frac{\ud^2 W}{\ud \omega \ud \Omega}=\frac{\alpha}{\pi^2}\left(
\frac{\theta}{\gamma^{-2}+\theta^2+\xi_1^2}-
\frac{\theta}{\gamma^{-2}+\theta^2+\xi_2^2}\right)^2 \ee
which holds for: $\gamma\gg 1, \quad \xi_1^2, \xi_2^2\ll 1, \quad \theta\ll 1$.
$\xi^2_i=\omega_{Pi}^2/\omega^2=1-\epsilon_i(\omega)$, where $\omega_{Pi}$ is 
the (electron) plasma frequency for the two media 
and $\alpha$ is the fine structure constant ($\alpha$=1/137).
The plasma frequency $\omega_P$ is a material property and can be
calculated as follows:
\be \omega_P = \sqrt{\frac{4\pi\alpha n_e}{m_e}} \approx 28.8\sqrt{\rho\frac{Z}{A}} 
\quad \mathrm{eV} \ee
where $n_e$ is the electron density of the medium and $m_e$ is the
electron mass. In the approximation $\rho$ is the density in $\mathrm{g/cm^3}$
and $\frac{Z}{A}$ is the average charge to mass ratio of the material. 
Typical values for plasma frequencies are $\omega_P^{CH_2}$=20.6 eV, 
$\omega_P^{Air}$=0.7 eV.

Since the emission angle $\theta$ of the TR is small ($\theta \simeq
\sqrt{\gamma^{-2}+\xi_2^2} \approx 1/\gamma$) one
usually integrates over the solid angle to obtain the differential energy 
spectrum:
\vspace{.1cm}
\be \left(\frac{\ud W}{\ud \omega}\right)_{interface}=\frac{\alpha}{\pi}\left(
\frac{\xi_1^2+\xi_2^2+2\gamma^{-2}}{\xi_1^2-\xi_2^2}\ln
\frac{\gamma^{-2}+\xi_1^2}{\gamma^{-2}+\xi_2^2}-2\right)\ee 

\begin{figure} %[hbt]
%\centering\mbox{\epsfig{file=./figs/trd-two.eps,width=0.5\textwidth}}
\centerline{\includegraphics[width=.5\textwidth]{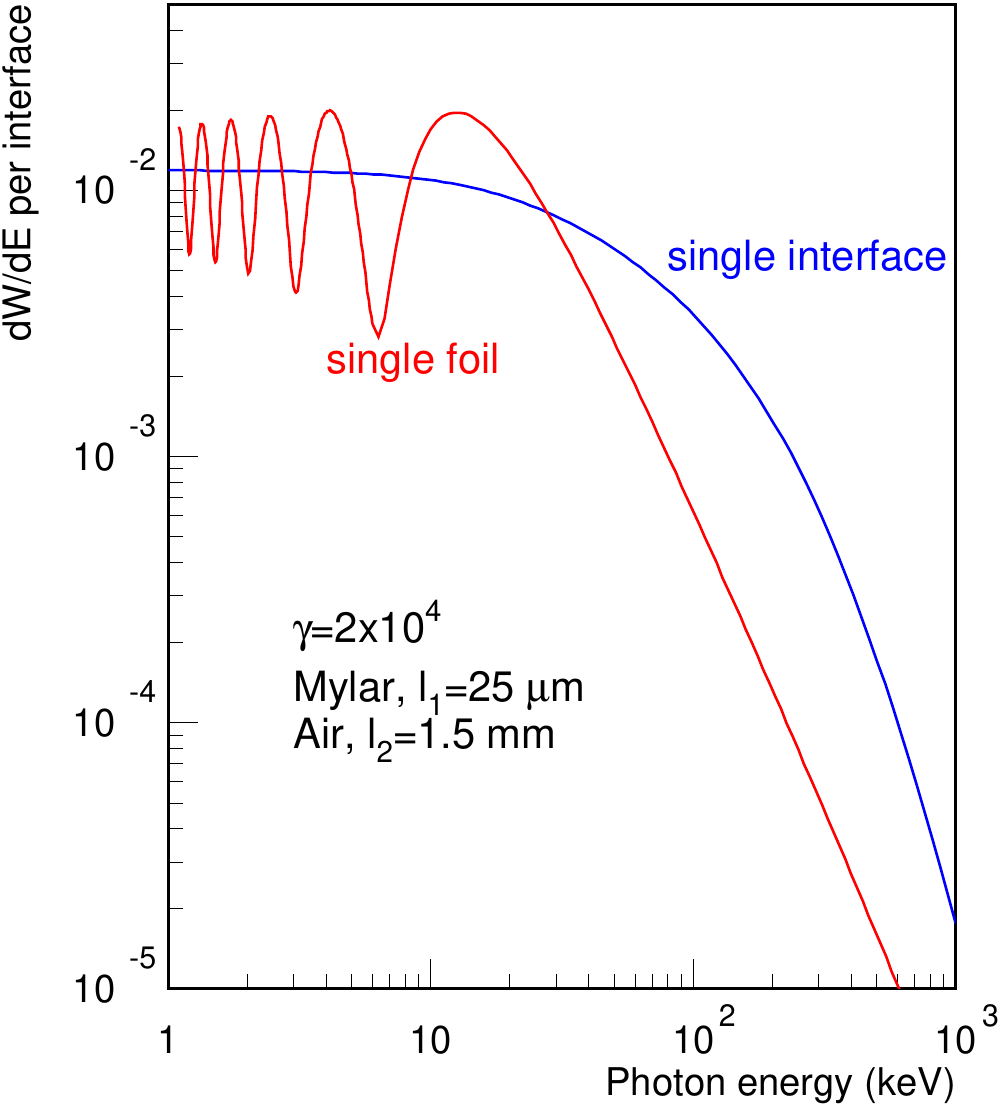}}
\caption{TR spectrum for single interface and single foil configurations.} 
\label{f:two} 
\end{figure}

A single foil has two interfaces to the surrounding medium at which
the index of refraction changes. Therefore, one needs to sum up the
contributions from both interfaces of the foil to the surrounding
medium. This leads to:
\vspace{.1cm} \be \left(\frac{\ud^2 W}{\ud \omega \ud
  \Omega}\right)_{foil}= \left(\frac{\ud^2 W}{\ud \omega \ud
  \Omega}\right)_{interface} \times 4\sin^2(\phi_1/2) \ee where
$4\sin^2(\phi_1/2)$ is the interference factor. The phase $\phi_1$ is
related to the formation length $Z_i$ (see below) and the thickness
$l_i$ of the respective medium,
i.e. $\phi_{i}\simeq(\gamma^{-2}+\theta^2+\xi_i^2)\omega
l_i/(2\beta\,c)$. Following the arguments in Ref.~\cite{che1} the average
amplitude modulation is $\langle 4\sin^2(\phi_1/2)\rangle\approx 2$. 
The above spectra are shown in Fig.~\ref{f:two} for one
interface of a single Mylar foil (25~$\mathrm{\mu m}$) in air
(using the same parameters as in Ref.~\cite{che1}).

Absorption of TR in the material of the radiator has not been considered
in the above. The effective TR yield, measured at the exit of the radiator,
is strongly suppressed by absorption for energies below a few keV \cite{che1}
(see also \cite{pdg}), see Fig.~\ref{f:tr_dep} below.

\subsection{TR production in regular multiple foil radiators}

As shown above the emission probability for a TR photon in the plateau region
is of order $\pi/\alpha$ per interface. For this to lead to a
significant particle discrimination one needs to realize many of
theses interfaces in a single radiator.
For a stack of $N_f$ foils of thickness $l_1$, separated by
a medium (usually a gas) of thickness $l_2$, the double differential
energy spectrum is:
\vspace{.1cm}
\be \left(\frac{\ud^2 W}{\ud \omega \ud \Omega}\right)_{stack}=
\left(\frac{\ud^2 W}{\ud \omega \ud \Omega}\right)_{foil}
\times \exp\left(\frac{1-N_f}{2}\sigma\right)
\frac{\sin^2(N_f\phi_{12}/2)+\sinh^2(N_f\sigma/4)}
{\sin^2(\phi_{12}/2)+\sinh^2(\sigma/4)} \ee
%\vspace{.1cm}
where $\phi_{12}=\phi_1+\phi_2$ is the phase retardation, with
$\phi_{i}\simeq(\gamma^{-2}+\theta^2+\xi_i^2)\omega l_i/2$,  
and $\sigma=\sigma_1+\sigma_2$ is the absorption cross section
for the radiator materials (foil + gas).
Due to the large absorption cross section below a few keV,
low-energy TR photons are mostly absorbed by the radiator itself.

The TR produced by a multi-foil radiator can be characterized by the following
qualitative features:

\clearpage

\begin{itemize}
\item One can define the so-called "formation zone" $Z_i$ 
\vspace{0.1cm}
\be 
Z_i=\frac{1}{\gamma^{-2}+\xi_i^2}\frac{2\,\beta c}{\omega}.
\ee
This can be interpreted as the distance beyond which the electromagnetic field
of the charged particle has readjusted and the emitted photon is separated 
from the field of the parent particle. The formation zone depends on the 
charged particle's $\gamma$, on the TR photon energy and is of the order of 
a few tens of microns for the foil and a few hundreds of microns for air \cite{dol}. 
The yield is suppressed if $l_i\ll Z_i$, which is referred to as the 
{\em formation zone effect}. \\
For constructive interference one gets:
\be \left(\frac{\ud^2 W}{\ud \omega \ud \Omega}\right)_{foil}=2\times
\left(\frac{\ud^2 W}{\ud \omega \ud \Omega}\right)_{interface} ; \quad
\left(\frac{\ud^2 W}{\ud \omega \ud \Omega}\right)_{stack}=N_f\times
\left(\frac{\ud^2 W}{\ud \omega \ud \Omega}\right)_{foil}\,\,. \ee

\item The TR spectrum has its most relevant maximum at
\vspace{0.1cm}
\be 
\omega_{max}=\frac{l_1\omega_{P1}^2}{2\pi \beta c} \label{eq:omega_max},
\ee  
which can be used to ``tune'' the TRD to the most relevant absorption
cross section of the detector
by varying the material and thickness of the radiator foils.

\item For $l_2/l_1\gg 1$ the TR spectrum is mainly determined by the single
foil interference.

\item The multiple foil interference governs the saturation at high
$\gamma$, above a value of
\be \gamma_s=\frac{1}{4\pi\beta c}\left[(l_1+l_2)\omega_{P1}+\frac{1}{\omega_{P1}}
(l_1\omega^2_{P1}+l_2\omega^2_{P2})\right].\ee
\end{itemize}

\noindent

\subsection{TR production in irregular radiators}

\begin{figure}[hbt]
\centerline{\includegraphics[width=.95\textwidth]{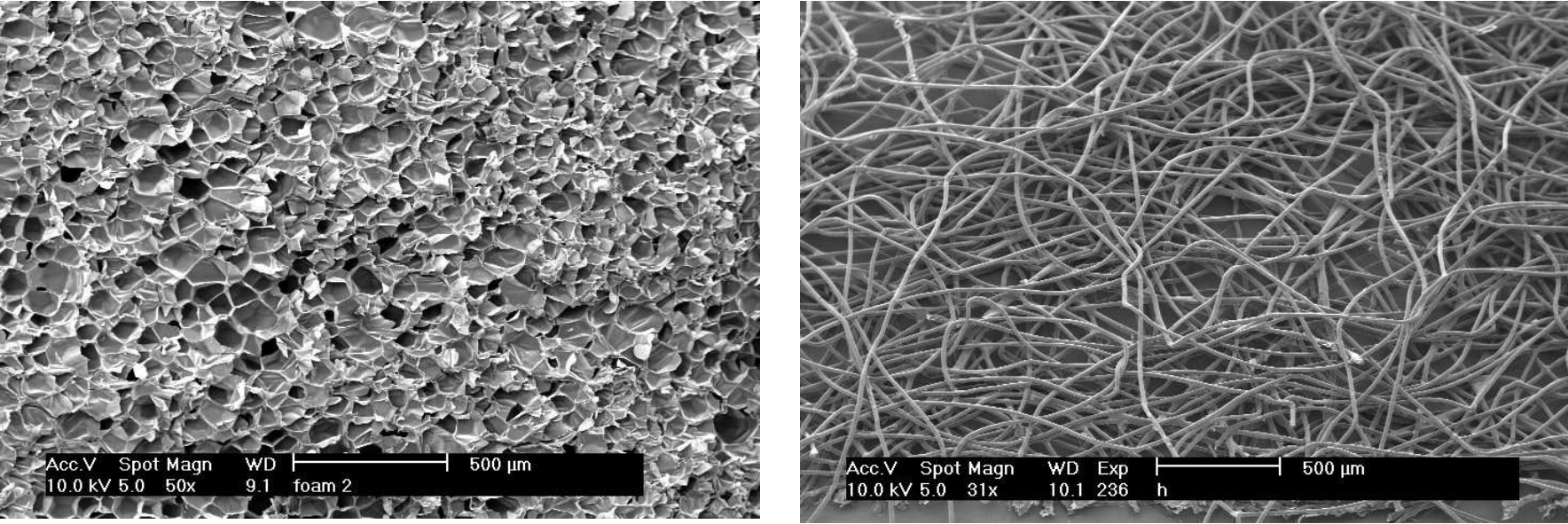}}
\caption{Electron microscope images of a polymethacrylimide foam
  (Rohacell HF71)(left) and a typical polypropylene fiber radiator
  (average diameter $\approx$ 25~$\mu$m) (right) \cite{alice-tr}.}
\label{f:rads} 
\end{figure}

In general, TR generated by irregular radiators can be calculated
following prescriptions discussed in Ref.~\cite{gar}. However, for all
practical purposes this procedure is limited to the treatment of
irregularities in the materials and tolerances from the fabrication of
otherwise regularly spaced radiators. For materials like foam or
fibers (used e.g. by HERMES, ATLAS in the central barrel, ALICE and AMS) 
as shown in Fig.~\ref{f:rads} this procedure is impractical. Here, the measured
response is simulated in terms of a regularly spaced radiator with
comparable performance \cite{alice-tr} or by applying an overall efficiency
factor~\cite{ege}.

\begin{figure}[htb]
\centerline{\includegraphics[width=.6\textwidth]{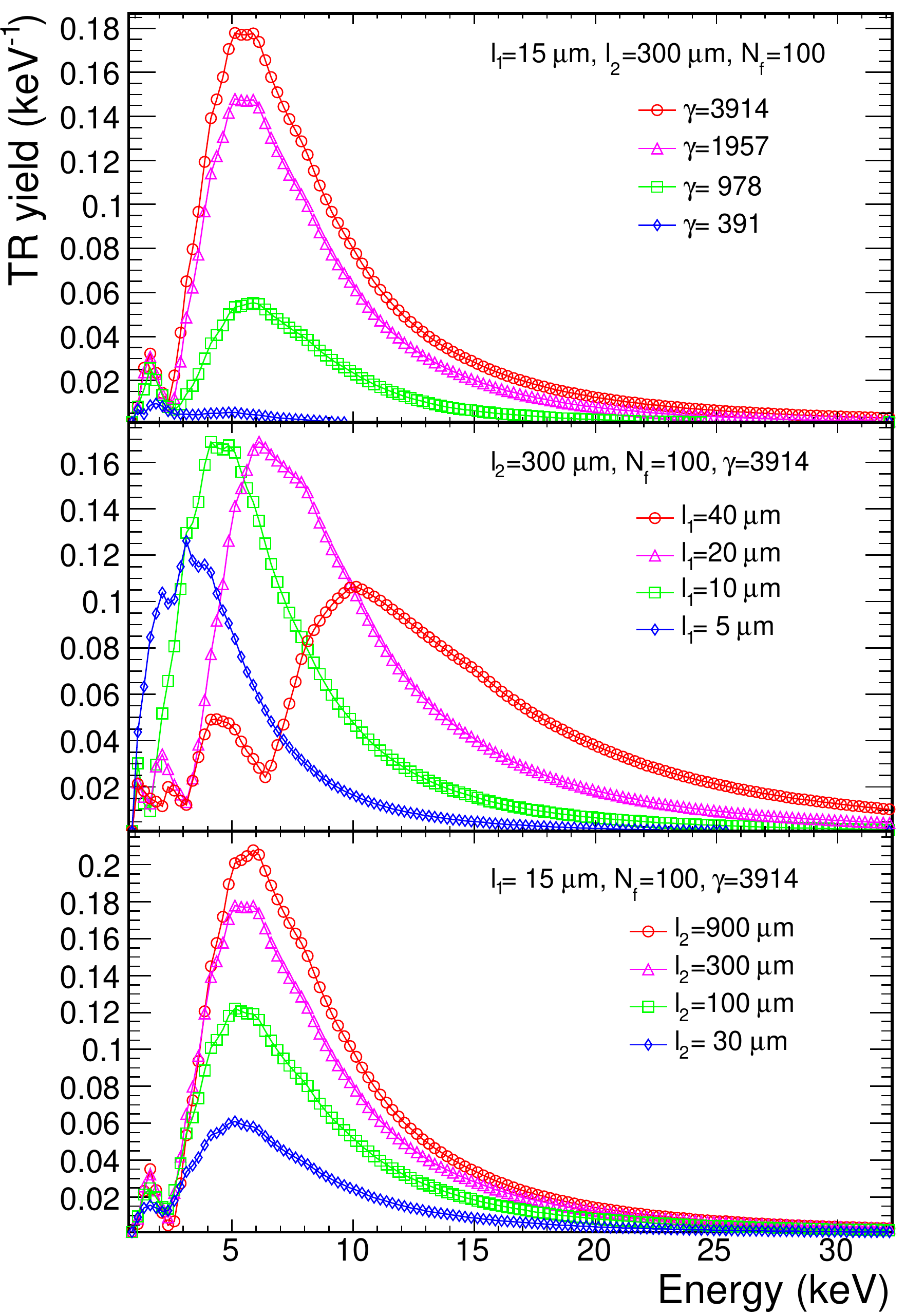}}
\caption{TR production as a function of: the Lorentz
  factor $\gamma$ (upper panel, corresponding to an electron momentum of 0.2, 
0.5, 1 and 2 GeV/c), foil thickness $l_1$ (middle panel) 
 and foil spacing $l_2$ (lower panel).}
\label{f:tr_dep} 
\end{figure}

\subsection{Basic features of TR production}

In Ref.~\cite{fab} TR has been studied for different (foil) radiator
configurations. 
The interference pattern discussed above has been demonstrated by 
Cherry et al.~\cite{che1} and by Fabjan and Struczinkski~\cite{fab}, who also
verified the expected dependence of the highest energy interference
maximum on the foil thickness, Eq.~\ref{eq:omega_max}.
In Ref.~\cite{fab} a slightly simpler expression for the TR production
has been proposed:

%\vspace{.1cm}
\be \frac{\ud W}{\ud \omega}=\frac{4\alpha}{\sigma(\kappa+1)}(1-\exp(-N_f\sigma))
\times \sum_n\theta_n\left(\frac{1}{\rho_1+\theta_n}-\frac{1}{\rho_2+\theta_n}
\right)^2 [1-\cos(\rho_1+\theta_n)] \label{tr1} \ee
where: 
\be\rho_i=\omega l_1/2\beta c(\gamma^{-2}+\xi_i^2), \quad \kappa=l_2/l_1,
\quad \theta_n=\frac{2\pi n-(\rho_1+\kappa\rho_2)}{1+\kappa}>0 \label{tr2}\ee

In the following we utilize this formula to show how the TR yield 
and spectrum (at the exit of the radiator) depend on the Lorentz factor 
$\gamma$ of the incident charged particle, as well as on the foil 
thickness ($l_1$) and spacing ($l_2$) for a regular radiator of $N_f$=100 foils.
These basic features of TR production are illustrated in Fig.~\ref{f:tr_dep}.
The threshold-like behavior of TR production as a function of $\gamma$
is evident, with the onset of TR production around $\gamma\simeq$1000.
The yield saturates quickly with $l_1$ (formation length for 
CH$_2$ is about 7~$\mu$m), the average TR energy is proportional to $l_1$, 
Eq.~\ref{eq:omega_max}.
Taking into account absorption of TR photons in the foils leads to an optimum 
of foil thickness in the range 15-20 $\mu$m (dependent also on
the thickness of the detector).
The TR yield is proportional to $l_2$ for gap values of a few hundred $\mu$m,
saturating slowly with $l_2$, as the formation length for air 
is about 700~$\mu$m; the spectrum is slightly harder for larger 
gap values.

Due to the dependence of TR on $\gamma$, it is evident that
there is a wide momentum range (1--100 GeV/c) where electrons
(resp. positrons) are the only particles producing transition radiation. 
Kaons can also be separated from pions on the basis of TR in a certain 
momentum range (roughly 200--700 GeV/c) \cite{errede} and 
$\pi/\Sigma$ identification in a hyperon beam has been done 
as well \cite{wa89}.

\section{From TR to TRD}\label{sect:det}

Having introduced the main features of TR production above, we shall
now focus on its usage for particle identification in high-energy 
nuclear and (astro-)particle experiments. We outline the main characteristics,
design considerations and optimization for a TRD, based on simulations.

\subsection{TR detection}

\begin{figure}[htb]
\centering\includegraphics[width=.56\textwidth]{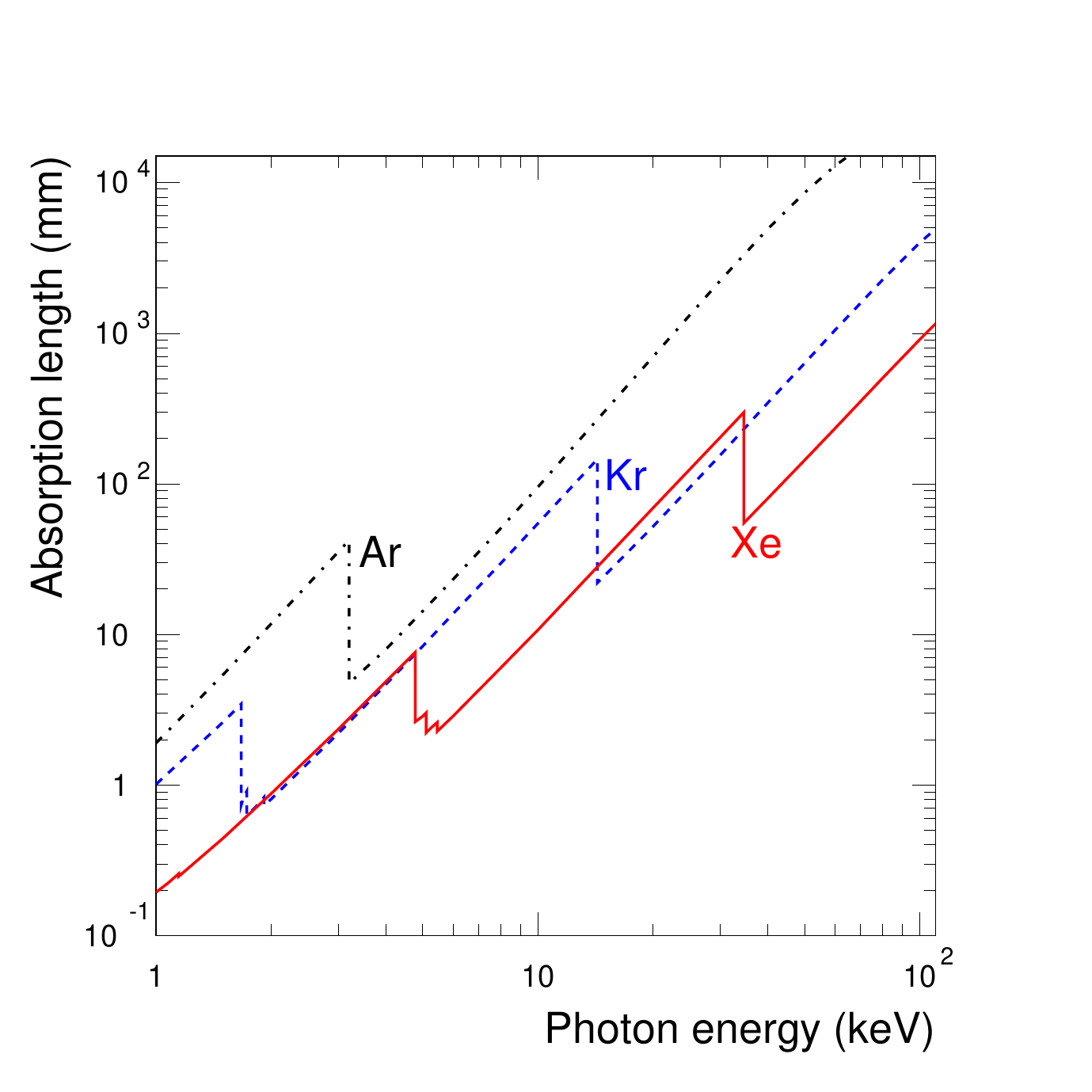}
\caption{Absorption length as a function of energy for X-rays in argon, 
krypton and xenon. The data was taken from the NIST database~\cite{nist}.}
\label{f:tr_gas} 
\end{figure}

An obvious choice to detect transition radiation is a gaseous detector.
A proposal to use silicon detectors in a TRD has also been put forward
\cite{sitrd} and TR detection with crystals has been proposed too 
\cite{access}, see below for more details.
Affordability for large-area coverage, usually needed in (accelerator)
experiments, is a major criterion. In addition, a lightweight construction
make gaseous detectors a widespread solution for TRDs.
Most of the TRD implementations are based on multiwire drift chambers, 
but straw tubes have been used too, for example in the NOMAD \cite{nom1}, 
HERA-B \cite{herab}
ATLAS \cite{atl}, PAMELA \cite{pam} and AMS \cite{ams} detectors.
We will describe the detector realization in the examples covered in 
Section~\ref{sect:modern_trds}.
See e.g. \cite{blum} for all the important details concerning drift
chambers principles and operation.
For gaseous detectors we present the absorption length \cite{nist} vs. TR
energy in Fig.~\ref{f:tr_gas} for Ar, Kr and Xe.
Obviously, the best detection efficiency is reached using the heaviest 
gas, Xe, which has an absorption length around 10 mm for ``typical'' 
TR photon energies in the range of 3-15 keV (produced by a radiator of 
typical characteristics, $l_1$=10-20 $\mu$m, $l_2$=100-300 $\mu$m, see 
Fig.~\ref{f:tr_dep}).
The electron identification is further enhanced by the ``favorable'' ionization
energy loss, $\ud E/\ud x$ in Xe, which has the highest value of the Fermi 
plateau of all noble gases.

\begin{figure}[htb]
\centering\includegraphics[width=.5\textwidth]{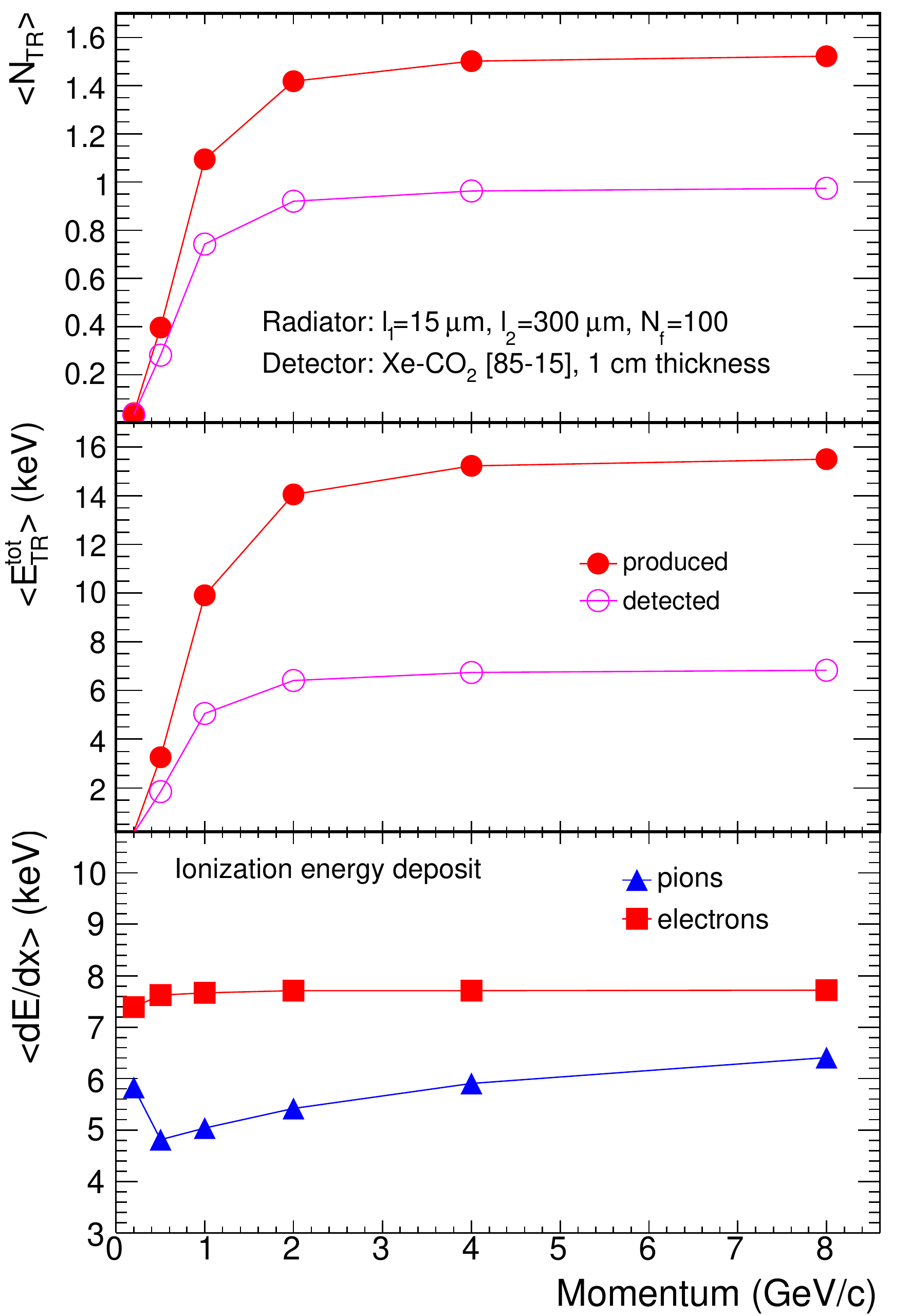}
\caption{Produced and detected average number of TR photons, $<N_{TR}>$
(upper panel), and total TR energy, $<E_{TR}^{tot}>$ (middle panel), as 
a function of electron momentum. In the lower panel we show for comparison 
the average ionization energy deposit, $<\ud E/\ud x>$, for pions and electrons.}
\label{f:tr_det} 
\end{figure}

In Fig.~\ref{f:tr_det} we consider a detector with a gas volume of
1~cm thickness and show its TR detection capability
as a function of momentum. On average, about 2/3 of the
number of produced TR photons (employing a radiator with $l_1$=15 $\mu$m, 
$l_2$=300 $\mu$m, $N_f$=100, which will be our baseline choice in the 
following) are detected in such a detector, filled with a mixture 
Xe-CO$_2$ [85-15]. 
About half the total produced TR energy $E_{TR}^{tot}$, which is 
the sum over all detected TR photons for an electron of a given momentum, 
is detected.
For the chosen configuration, on average the signal from TR is comparable to the 
ionization energy deposit, $<\ud E/\ud x>$, also shown in Fig.~\ref{f:tr_det}.

\subsection{Basic performance characteristics of a TRD}

\begin{figure}[hbt]
\begin{tabular}{lcr}
\begin{minipage}[t]{0.48\textwidth}
\centering\includegraphics[width=1.05\textwidth]{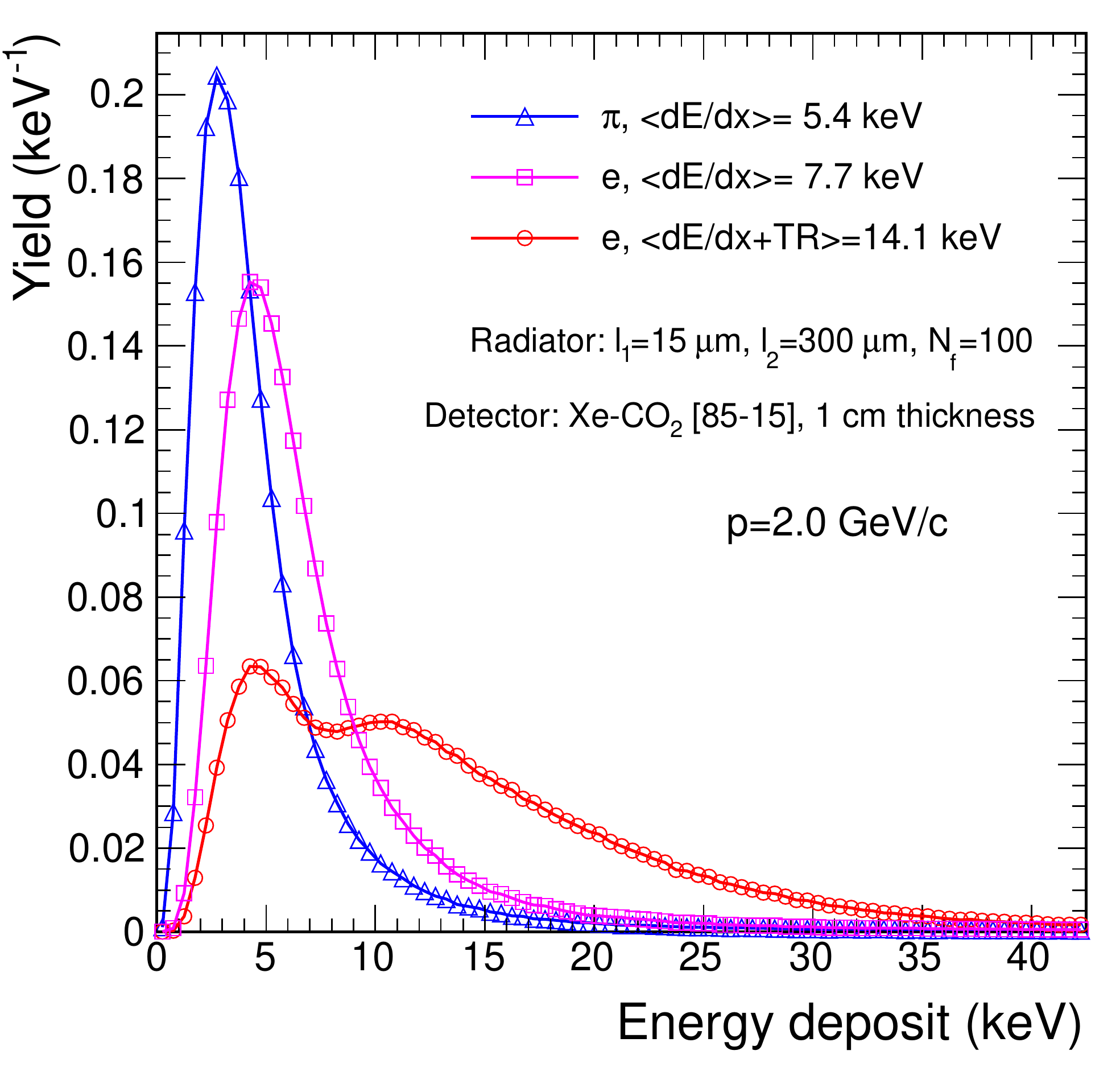}
\end{minipage} &
\begin{minipage}[t]{0.48\textwidth}
\centering\includegraphics[width=1.05\textwidth]{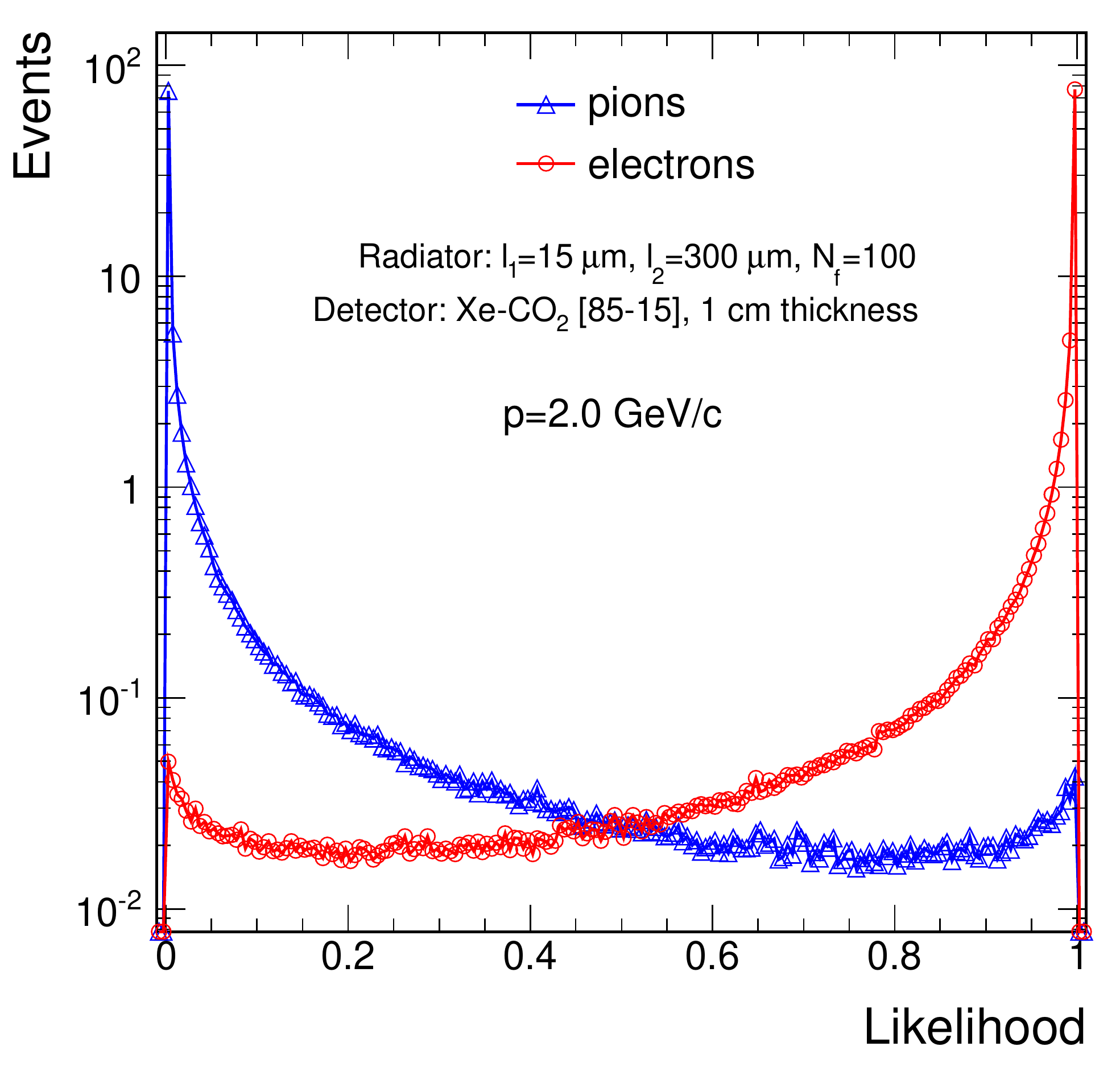}
\end{minipage}
\end{tabular}
\caption{Spectra of the total energy deposited in one layer of a TRD 
for pions and electrons (left panel) and electron likelihood distributions 
constructed from them for six layers (left panel).
}
\label{f:like}
\end{figure}

It is important to emphasize that, due to the very small TR emission angle, 
the TR signal generated in a detector is overlapping with the ionization due 
to the specific energy loss $dE/dx$ and a knowledge (and proper simulation)
of dE/dx~\cite{alice-dedx} (see also Section~\ref{sect:alice}) is a necessity
for the ultimate understanding and modeling of any TRD.
The energy deposit spectra of pions and electrons in a Xe-based detector 
are presented in Fig.~\ref{f:like} (left panel). For pions it represents 
the energy loss in the gas and is close to a Landau distribution. 
For electrons, it is the sum of the ionization energy loss and 
the signal produced by the absorption of the TR photons.

It is evident that due to the large tails in the energy loss
spectrum for pions, the detector has to have many layers. In case the
full charge signal is available the 
discrimination is done using either a normal mean, a truncated mean 
(discarding the highest measured value of the detector sets) 
or, preferably, a likelihood method \cite{bun,zeus,nom2}.

Likelihood distributions are constructed from the measured spectra of
identified particles. Taking these spectra (for each layer) as probability 
distributions for electrons ($P(E_i|e)$) and pions ($P(E_i|\pi)$ ) to
produce a signal of magnitude $E_i$,  one constructs the 
{\em likelihood} (to be an electron) as \cite{ap0}:
\be 
\mathrm{likelihood}=\frac{P_e}{P_e+P_\pi}, \quad
P_e=\prod_{i=1}^NP(E_i|e), \quad
P_\pi=\prod_{i=1}^NP(E_i|\pi) 
\label{e:l1}
\ee
or, equivalently (also called log-likelihood) \cite{nom2}: 
\be
\mathrm{likelihood}=\sum_{i=1}^N\log\frac{P(E_i|e)}{P(E_i|\pi)} 
\label{e:l2} 
\ee 
where the product (sum) runs over the number of detector layers.  
The likelihood defined by Eq.~\ref{e:l1} is shown in Fig.~\ref{f:like}
(right panel) for pions and electrons, for a likelihood 
derived from the integrated charge signal.
The electron identification performance of a TRD is quantified in terms
of the {\em pion efficiency} at a given electron efficiency and is the 
fraction of pions wrongly identified as electrons.

\begin{figure}[htb]
\begin{tabular}{cc}
\begin{minipage}{.48\textwidth}
\centering\includegraphics[width=1.05\textwidth]{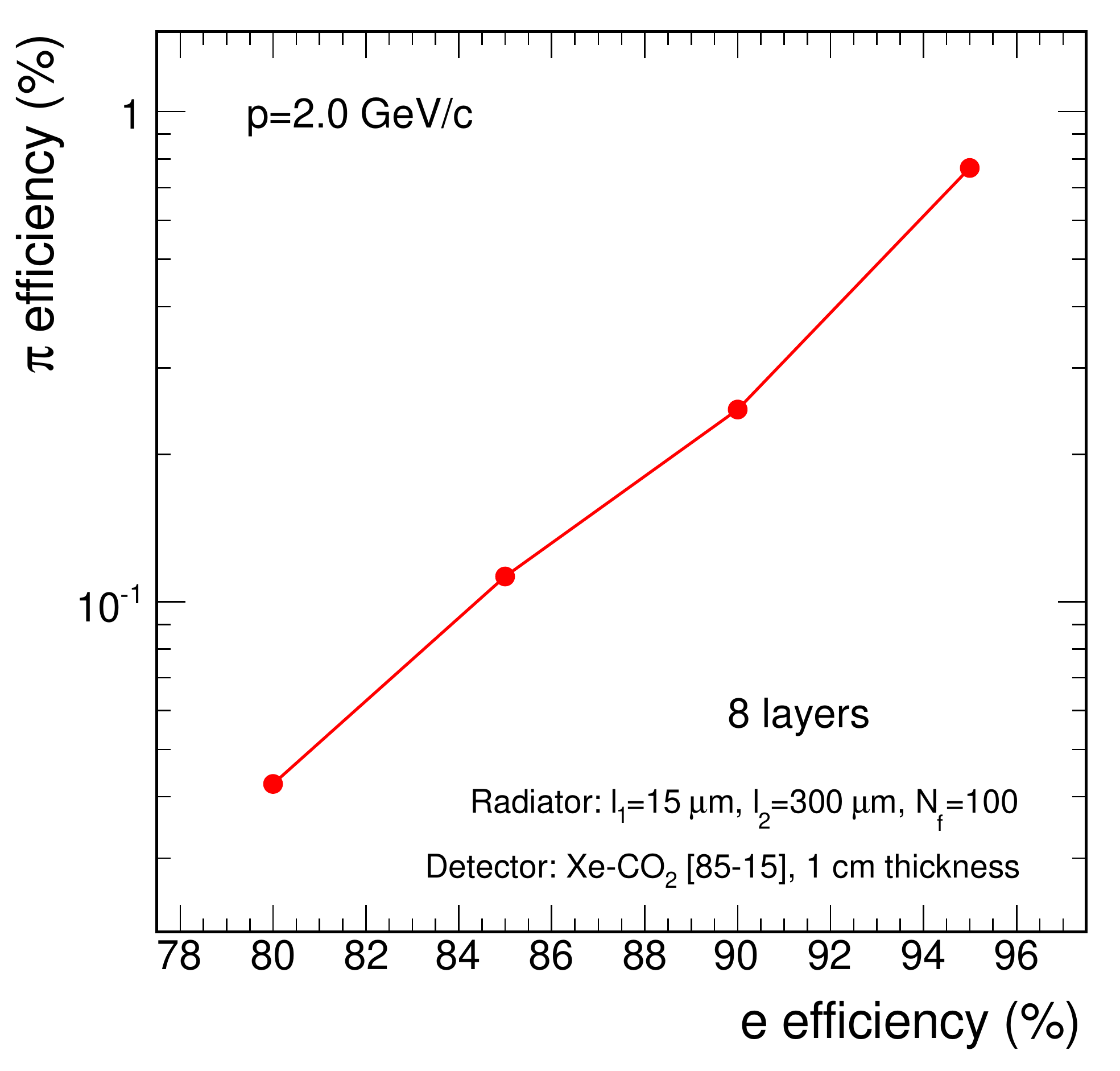}
\end{minipage} & \begin{minipage}{.48\textwidth}
\centering\includegraphics[width=1.05\textwidth]{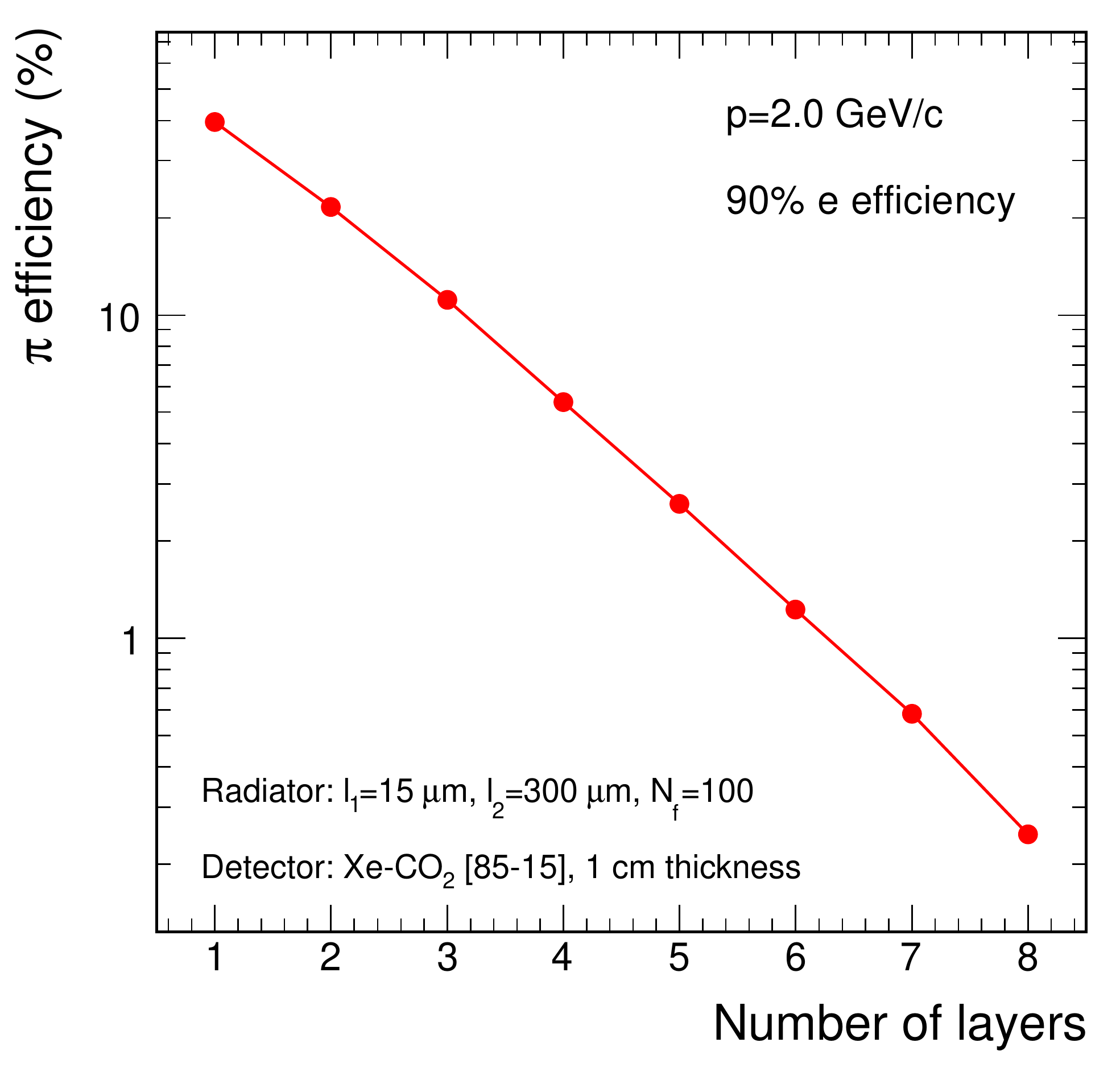}
\end{minipage} \end{tabular}
\caption{Pion efficiency as a function of the electron efficiency (left panel) 
and as a function of the number of layers (right panel).} \label{f:pieff}
\end{figure}

In Fig.~\ref{f:pieff} we show the calculated pion efficiency as a function of 
electron efficiency and as a function of the number of layers for a momentum 
of 2 GeV/c. The pion efficiency depends strongly on the electron efficiency,
which is a parameter that can be adjusted at the stage of the data analysis.
It is chosen such that the best compromise between electron efficiency
and purity is reached.
Usually 90\% electron efficiency is the default value used to quote a TRD 
rejection power and we will use this value throughout the paper.
Even more crucial for the design of the detector is the dependence on the number 
of layers, which has a lot of implications, see below.

\begin{figure}[hbt]
\centering\includegraphics[width=.55\textwidth]{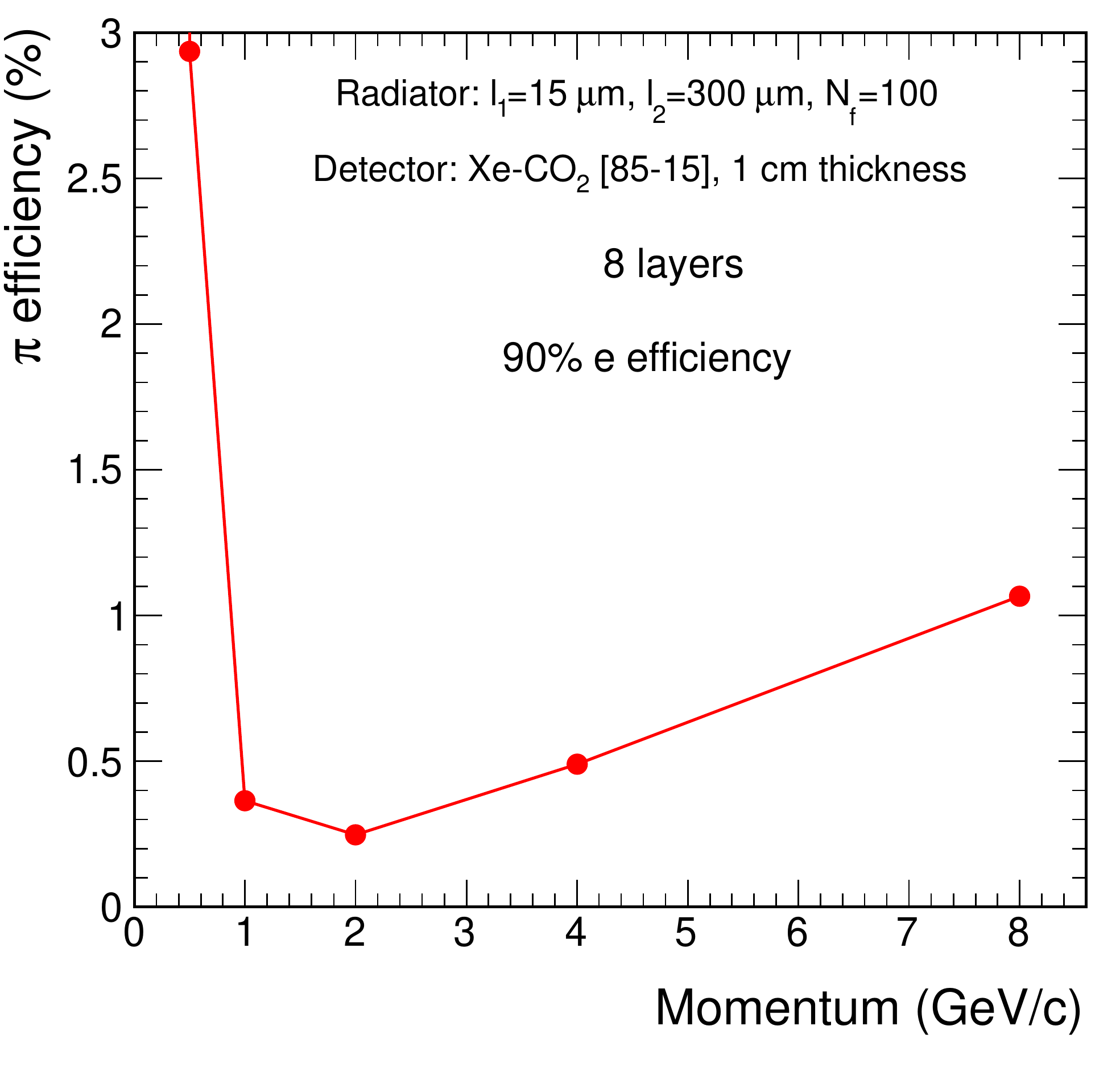}
\caption{Pion efficiency as a function of momentum.}
\label{f:pieff-p} 
\end{figure}

The dependence of the pion efficiency on momentum is shown in Fig.~\ref{f:pieff-p}.
The rejection power improves dramatically between 0.5 and 1~GeV/c and slightly 
up to 2~GeV/c as a result of the onset of TR production in this momentum range. 
Beyond 2~GeV/c TR yield saturates and the rejection power decreases as a 
consequence of the relativistic rise of the ionization energy loss of pions. 
While in detail this behavior depends on the specific choice of the radiator, 
it is a generic feature of the TRD pion rejection capability.

\subsection{TRD design considerations}

Considering the above mentioned properties of TR generation and
absorption, it is evident that a TRD requires careful optimization
concerning the following aspects governing the radiator and detector
design:
\subsubsection{ Radiator}
\begin{itemize}
\item 
{\em Type:} regular (foils) vs. random (foams, fibers). Owing to the
interference effects mentioned above foil radiators were 
shown to produce more TR photons than foams \cite{ed1,alice-tr} at comparable
density. However, a number of fiber radiators have shown comparable performance 
to foil radiators~\cite{fab,bun,but,wat,zeus,hol,gra,ktev,her,alice-tr}.
\item 
{\em Material:} since many interfaces are necessary the foil material
itself needs to have an as low X-ray absorption coefficient
as possible. Li~\cite{cob}, Be, polypropylene - CH$_2$~\cite{ed2,nom1}, 
and mylar~\cite{che1} have been used.
The same argument holds for the gas inside the gaps. He~\cite{h1}
would be the preferred choice for its low absorption 
cross section. However, for practical reasons (like special containment 
vessels etc.) most commonly air~\cite{ed2}, N$_2$~\cite{nom1},
or CO$_2$~\cite{na31} are being used.
\item 
{\em Configuration:} as shown above, the foil and gap thicknesses ($l_1$, $l_2$)
and number of foils ($N_f$) determine the TR production yield and spectrum.
The foil thickness can be matched to the detector thickness.
The thickness of the gap between foils shall be ideally around 1~mm 
(see Fig.~\ref{f:tr_dep}). However, for a given total radiator thickness, a compromise
needs to be found between the total number of foils per radiator and the foil gap. 

\end{itemize}

\subsubsection{Detector}
\begin{itemize}
\item
{\em Gas:} type and thickness. As discussed above, Xe is the best choice for 
the main gas, while the choice of the quencher is broad and dictated by different 
arguments. For instance, methane, CH$_4$, is an effective quencher but
its flammability forbids its usage in collider experiments.
Nowadays CO$_2$ is widely used.
Ref.~\cite{dol,dol1} provide a comprehensive discussion on gas properties
of relevance to TRDs. 
Cleanliness of the gas is important, both in view of signal collection
(avoiding electron attachment, in particular for drift chambers \cite{alice-att})
and to avoid aging~\cite{ageing}.
\item
{\em Read-out and signal processing:} the different options include integral 
charge detection~\cite{na31,wa89,nom1,her}, cluster counting~\cite{fab2,zeus,dol,atl}
or a time-resolved amplitude readout using FADCs~\cite{wat,hol,gra,ans,zeus,d0,h1}.
The latter allows to use both integral charge and cluster counting for 
TR recognition.
\end{itemize}

\begin{figure}[hbt]
\centering\includegraphics[width=.55\textwidth]{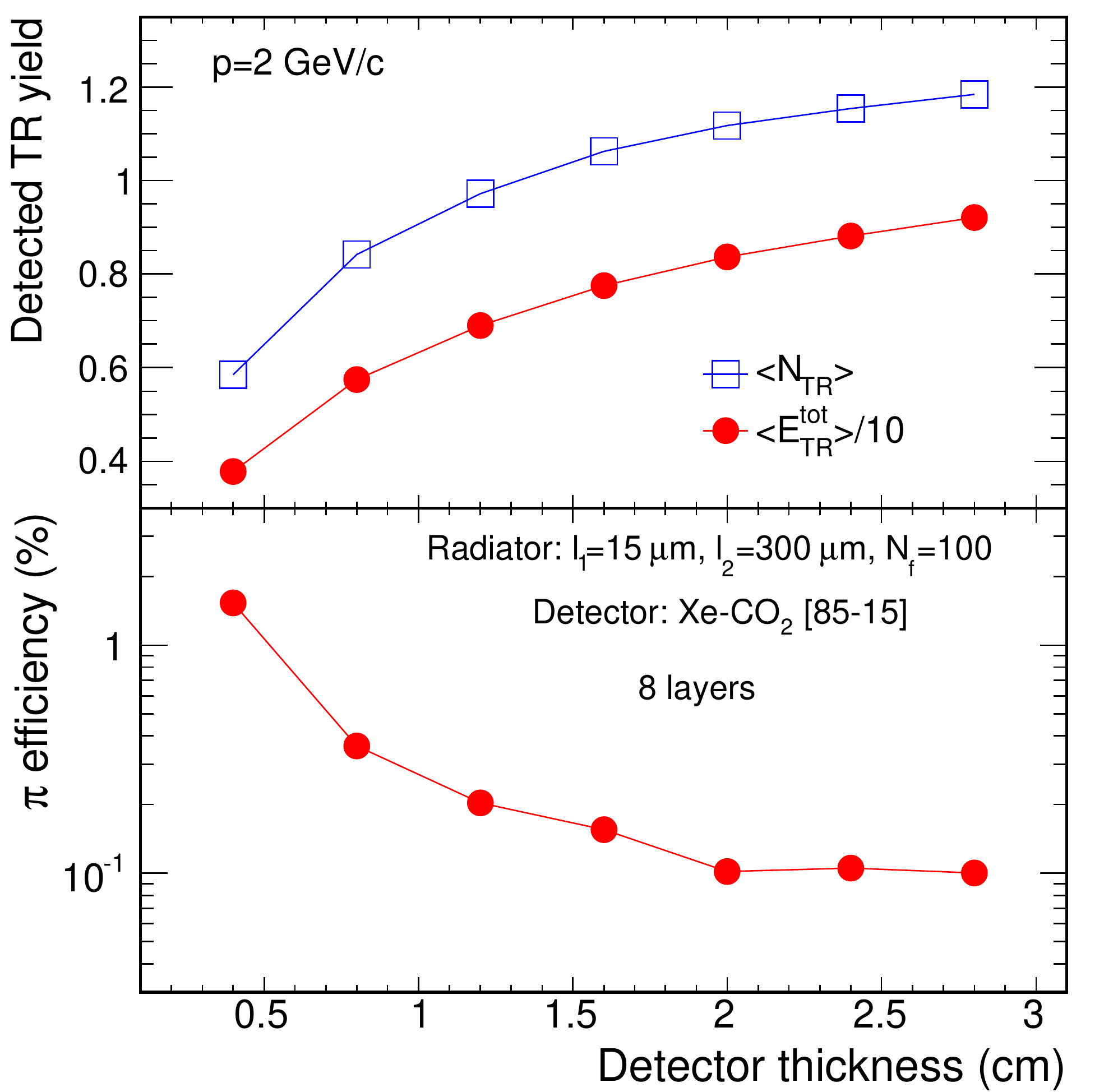}
\caption{Detected TR yield (upper panel, note the factor of 10 scale used for 
plotting of the detected total TR energy, $<E_{TR}^{tot}>$) and pion efficiency 
for 8 layers as a function of the single (layer) detector thickness. }
\label{f:pieff-x} 
\end{figure}

Obviously, the depth of the detector should be large enough to ensure efficient 
X-ray absorption, and, if included in the detector design, to allow for a better 
pion rejection by exploiting the position information of the clusters.
Given the cost of Xe, the total volume of the detector and gas system
also needs to be considered.
The dependence of the TRD performance on the thickness of the detection element
is shown in Fig.~\ref{f:pieff-x}. We consider 8 layers of a radiator and
a detector of variable thickness. On average a factor 2 more TR photons are
detected if one increases the thickness of the detector from 0.4~cm to 3~cm.
For the total detected TR energy the factor is slightly larger than 2 and the 
overall rejection power increases by more than an order of
magnitude. The most significant
improvement is seen up to a detector thickness of about 1.5~cm, with rather
marginal gain for an even thicker detector.

As shown above, the number of layers (providing independent signal values for 
a particle) is a crucial choice for a TRD.
Limitations arise, for large-scale detectors, due to budgetary constraints 
and sometimes even due to space availability in a detector setup. 
Another constraint can be the total amount of material a TRD represents: 
usually, $X/X_0\sim$10-20\% is achievable for a TRD. This usually tolerable 
for other detector systems (like time-of-flight or calorimeters) placed behind 
the TRD in an experiment.
For a given overall detector depth, there is a ``trade-off'' between 
the number of layers and the layer thickness. A thin layer design 
allows a faster detector readout, while a thick layer option, if speed
is not a crucial requirement, may be more advantageous for large track densities
requiring good readout granularity.
The two approaches, as implemented in the ATLAS \cite{atl} and ALICE \cite{ali},
respectively, are illustrated in detail in the following sections.

\begin{figure}[hbt]
\centering\includegraphics[width=.55\textwidth]{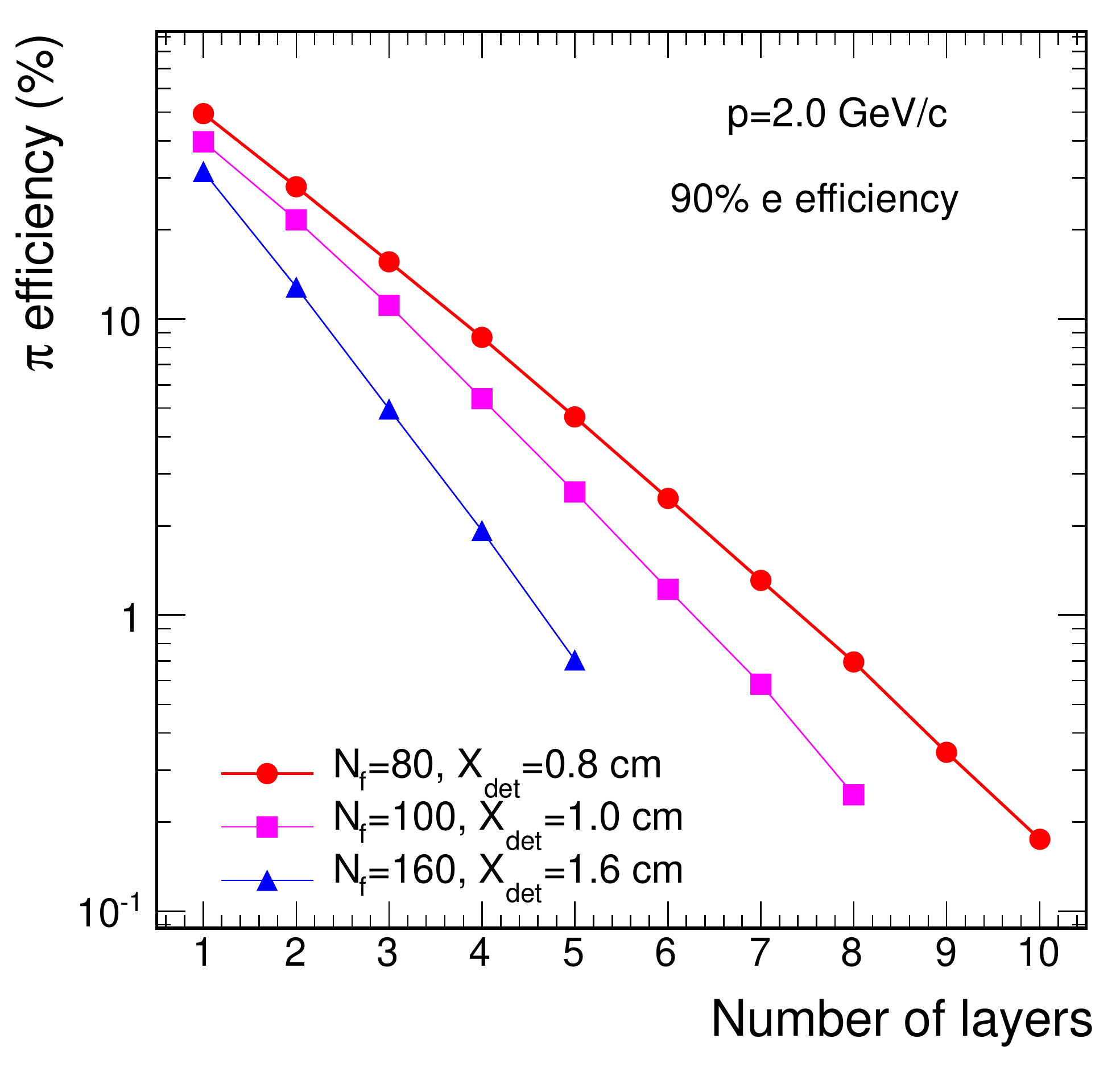}
\caption{Pion efficiency as a function of the number of layers for three
scenarios of detector granularity.}
\label{f:pieff-y} 
\end{figure}

The dependence of the pion efficiency on the number of layers (for 90\% electron 
efficiency) is shown in Fig.~\ref{f:pieff-y} under the constraint of a
constant overall radiation length of the entire TRD.

In estimating the pion efficiency, 
we have here focused on the total charge recorded in a detector layer.
Sometimes the total charge is obtained by integrating above a certain 
low value threshold (typically 10 times the noise level)~\cite{nom2}.
In drift chambers, the clusters are counted if they are above a high value 
threshold. Often, the threshold used is a variable one, increasing as a 
function of the drift time (``intelligent threshold'')~\cite{ed1}.
An improved version of this method is the time-over-threshold method 
(ToT)~\cite{tot}, which is applied also for thin detectors. 
The total charge in (large) clusters was also used~\cite{zeus}.
It was pointed out in Ref.~\cite{hol} that a bidimensional likelihood
on cluster position and integral charge improves the rejection by 
a factor of 2 compared to the likelihood based on the integral
charge. A somewhat smaller 
improvement with this method was recently measured~\cite{alice-trd2d}.
Neural networks were also proposed~\cite{bari-neural,alice-neural} 
as a powerful method, which can be used when the time-sampled signal 
information is available.
They can provide a significant improvement of the pion rejection~\cite{alice-ppr}
(see also below).

\begin{figure}[hbt]
%\centering\mbox{\epsfig{file=./figs/TRD_07.eps,width=0.6\textwidth}} 
\centering\includegraphics[width=.55\textwidth]{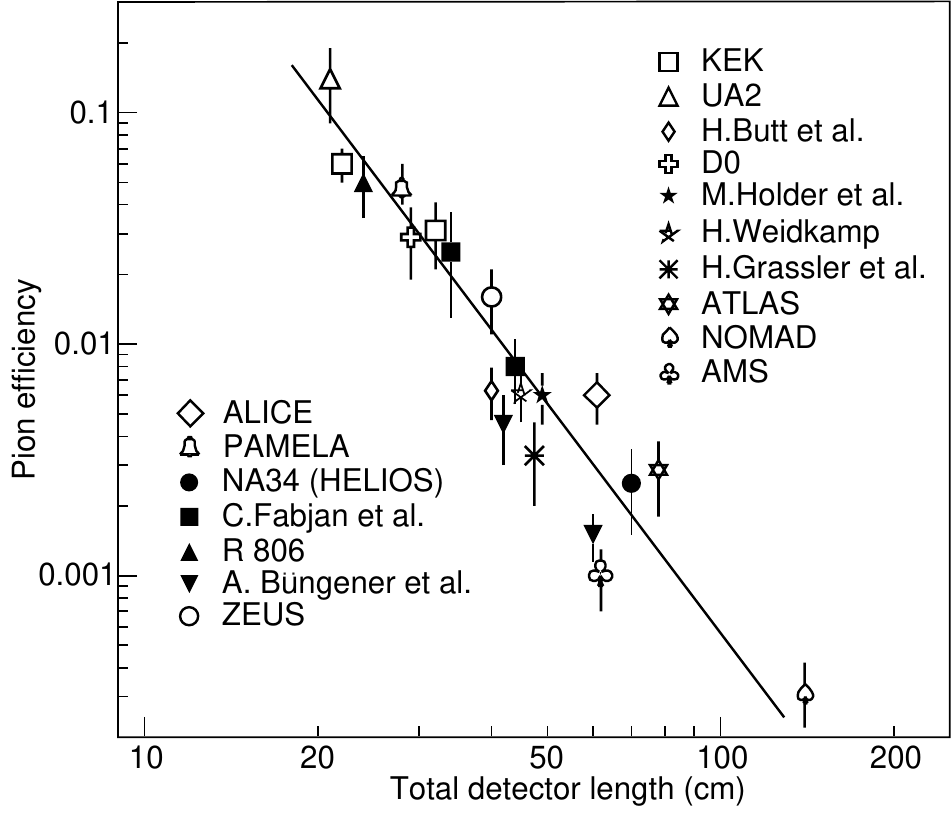}
\caption{TRD rejection power as function of the total length of the detector
for various high-energy (astro-)particle experiments (figure from \cite{pdg}).
The line is drawn to guide the eye.} 
\label{f:rejj}
\end{figure}

\subsubsection{Synopsis of TRDs used in different experiments}

\begin{sidewaystable} %[htb]
%\begin{table}[htb]
\centering
{
 \begin{tabular}{l|c|c|c|c|c|c|c} 
Experiment ~&Radiator (x,cm) &Detector (x,cm) &Area(m$^2$) / N &L (cm) 
& N. chan. & Method &~~ $\pi_{rej}$~~ \\ \hline
{\bf HELIOS}& foils (7) &Xe-C$_4$H$_{10}$ (1.8) & 4 / 8 & 70 &1744 &N &2000\\ 
\hline
{\bf H1} &foils (9.6) &Xe-He-C$_2$H$_6$ (6) &5.3 / 3 &60 & 1728 &FADC &10\\ \hline
{\bf NA31} & foils (21.7)&Xe-He-CH$_4$ (5)& 18 / 4&96 &384 &Q &70  \\ \hline
{\bf ZEUS} &fibres (7) &Xe-He-CH$_4$ (2.2) &12 / 4 &40 &2112 &FADC &100\\ \hline
{\bf D0}   & foils (6.5) &Xe-CH$_4$ (2.3) &11 / 3 & 33  &1536 &FADC &50 \\ \hline
{\bf NOMAD} &foils (8.3) &Xe-CO$_2$ (1.6) &73 / 9 &150 &1584 & Q  &1000 \\ \hline
{\bf HERMES} & fibres (6.4) &Xe-CH$_4$ (2.54) &28 / 6 & 60 &3072 & Q &1400 \\ 
\hline %.75x3.25x2
{\bf kTeV} & fibres (12) &Xe-CO$_2$ (2.9) & 39 / 8 &144 & $\sim$10 k &Q &250 
\\ \hline
{\bf PHENIX}& fibres (5) &Xe-CH$_4$ (1.8) &300 / 6  & 4 &43 k &FADC &$\sim$300 \\
\hline
{\bf PAMELA} & fibres (1.5) & Xe-CO$_2$ (0.4) & 0.7 / 9 & 28 & 964 & Q,N & 50 \\ 
\hline
{\bf AMS} & fibres (2) & Xe-CO$_2$ (0.6) & 30 / 20 & 55 & 5248 & Q & 1000 \\ 
\hline
{\bf ATLAS} & fo/fi (0.8) &Xe-CO$_2$-O$_2$ (0.4)& 1130 / 36 & 40-80 &351 k &
N,ToT &100 \\ \hline
%\hline
{\bf ALICE}& fi/foam (4.8) &Xe-CO$_2$(3.7)& 716 / 6 &52 &1.2 mil. &FADC &200 \\
\end{tabular}
}
\caption{TRD characteristics used in high-energy (astro-)physics experiments.}
\label{t:trd}
%\end{table}
\end{sidewaystable}

In Table~\ref{t:trd} we show the main characteristics and the
performance of TRDs used in various high-energy (astro-)particle experiments.

Fig. \ref{f:rejj} shows the rejection power of TRD vs. its total length
\cite{gra,dol}, as compiled recently in \cite{pdg}, for the configurations 
presented in the above table and for the ones in Ref. 
\cite{fab2,bun,ans,det,gra,but,wat,hol}. 
In most of the cases the momentum range was below 10 GeV/c.
There is a clear improvement of the rejection power when the overall length 
of the detector increases. 
But one should keep in mind that the performance of a TRD depends also on 
the configuration (``granularity'', as we discussed above) and signal processing
(most of the results have been obtained using the likelihood method).
No relevant difference is observed between charge (``Q'') and cluster counting 
(``N'') methods \cite{dol}.
It is also apparent that, within errors fiber radiators are comparable to 
foil radiators in terms of performance vs. length~\cite{zeus,hol}. 
However, scaled to the total radiation thickness, foil radiators are often
superior in terms of their TR yield~\cite{alice-tr}.

\subsection{Further developments}

\subsubsection{Heavy element detection}

A novel application of TRDs is in the detection of nuclei in cosmic rays of 
energies up to 100 TeV/nucleon~\cite{access}. This requires a different design 
as compared to TRDs discussed above in terms of radiators (material and geometry) 
as well as detectors. For space-borne experiments, additional
constraints specific to space instruments have to be considered.
The proposed concept of ACCESS~\cite{access} is different compared to
a ``classic'' TRD in the following aspects:

\begin{figure}[hbt]
\centering\includegraphics[width=.8\textwidth]{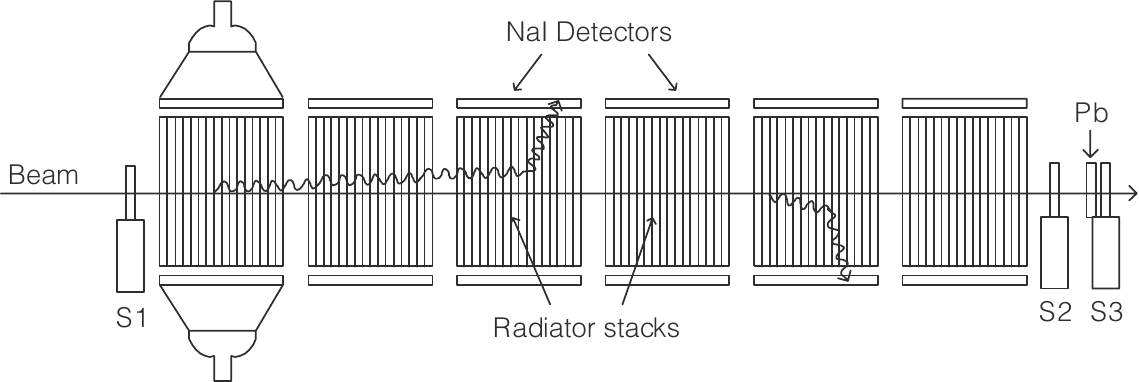}
\caption{Schematic layout for a test performed with a prototype for
  the ACCESS experiment (from~\cite{access}).}
\label{f:access}
\end{figure}

\begin{itemize}
\item
 The radiator is constructed from foils of 100-200 $\mu$m thickness, made of 
either mylar or Teflon. An aluminum honeycomb structure was tested too. The foil
spacing is around 3.5 mm. The resulting TR has a harder spectrum compared to 
the cases discussed above. The most probable value is in the range of 60-230~keV.
\item 
The detector employs NaI(Tl) crystals readout by photomultipliers via 
light guides, which collect the Compton scattered TR at 90$^\circ$
with respect to the 
charged particle incidence (which is perpendicular to the radiator) as
depicted in Fig.~\ref{f:access}.
\end{itemize}

\begin{figure}[hbt]
\centering\includegraphics[angle=-90,width=.95\textwidth]{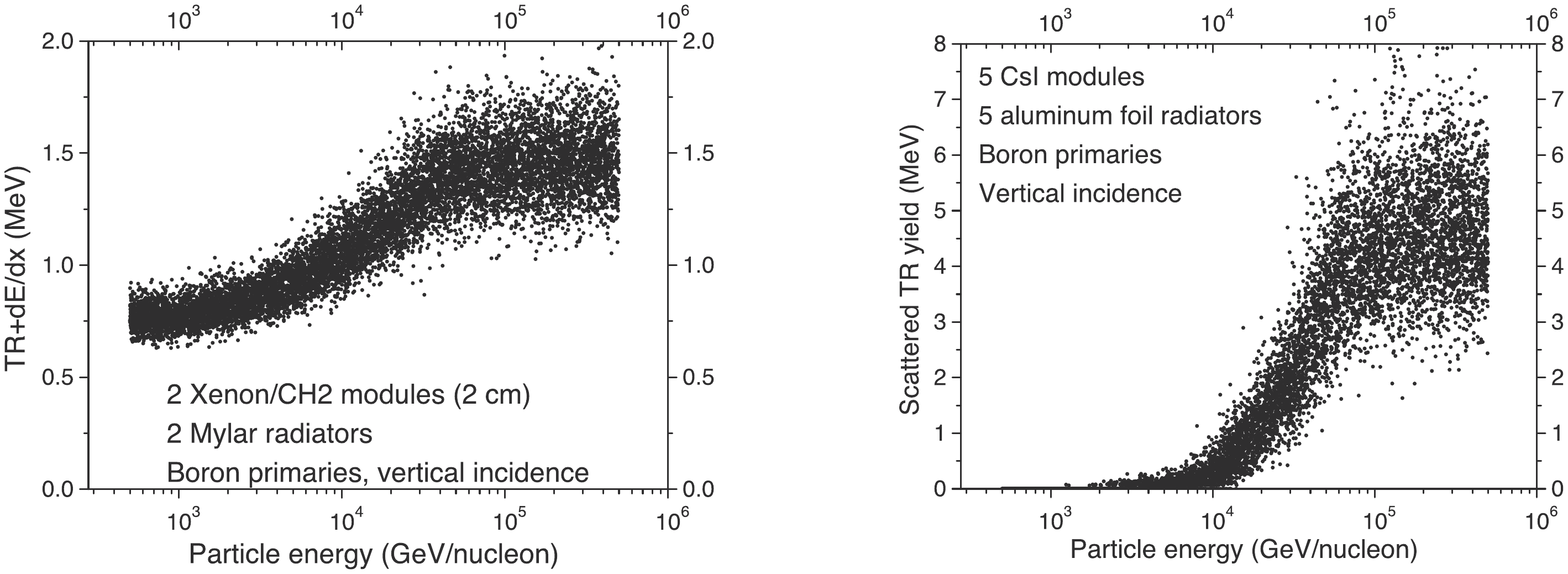}
\vspace{-2cm}
\caption{Response of mylar (left) and aluminum radiators (right) to
  boron nuclei, for details see~\cite{boron}.}
\label{f:boron}
\end{figure}

On average, up to about 0.03 photons are detected~\cite{access} from each 
radiator, depending on its type, for particles with $Z$=1 with Lorentz factors 
around $\gamma$=10$^5$. Short of beams of heavy elements with the
mentioned $\gamma$-factors, Fig.~\ref{f:boron} shows a comparison of an
aluminum based radiator (right panel) employing the above mentioned
Compton scattered X-rays ( from boron nuclei) detected with CsI-crystals
compared to a mylar radiator (left panel) read out via a 
Xe-based detector~\cite{boron}.

\subsubsection{Silicon-TRD}

\begin{figure}[hbt]
%\centering\mbox{\epsfig{file=./figs/sitrd.eps,width=0.8\textwidth}} 
\centering\includegraphics[width=.95\textwidth]{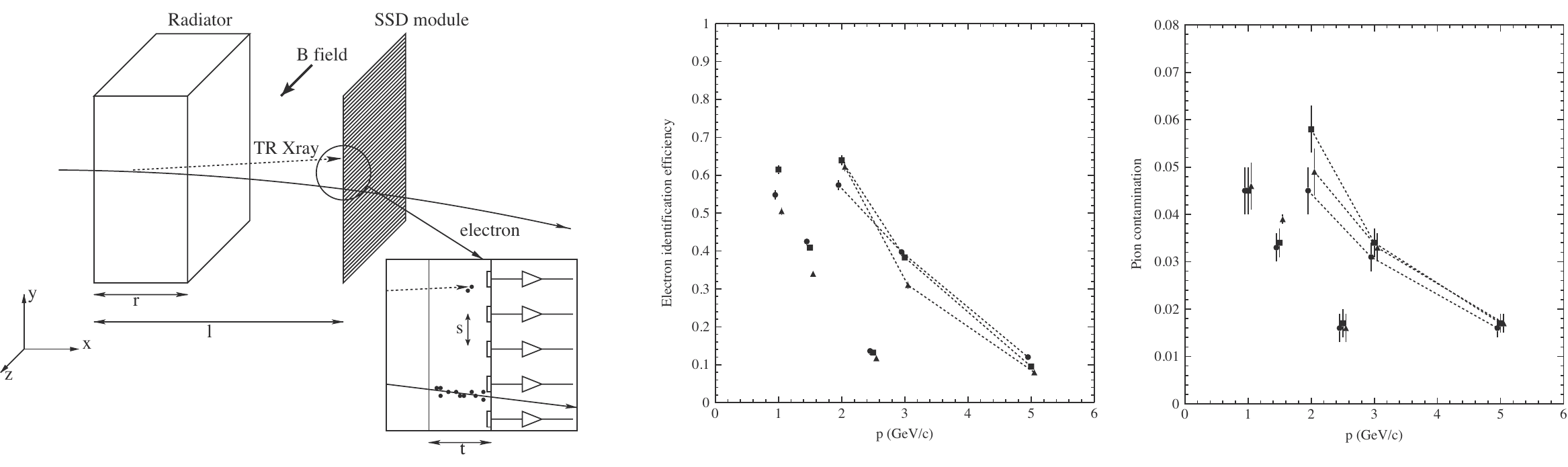}
\caption{Left panel: Schematic drawing of the operating principle of a
  Si-TRD; middle panel: Electron identification efficiency; right
  panel: pion contamination for three different SIGRAFIL C radiator
  thicknesses (15~mm (circles), 30~mm (squares), 45~mm (triangles))
  taken at magnetic fields of 0.44~T (dotted lines) and 0.87~T (dashed
  lines), from~\cite{sitrd}.}
\label{f:sitrd}
\end{figure}

An interesting new approach is based on TR detection using silicon strip 
detectors \cite{sitrd}. The method is based on the explicit separation
of the measurement of TR from the signal from ionization, exploiting the 
position resolution of the Si-strip detector and deflection of the charged 
particle in a magnetic field.
The prototype realization and performance of such a SiTRD was
demonstrated in \cite{sitrd}. A schematic drawing of the detection
principle is shown in the left panel of Fig.~\ref{f:sitrd}.

While pion rejection factors of about 20 (for momenta of 1-2 GeV/c)
to about 65 (at 5 GeV/c) were achieved, the electron efficiency
actually dropped with increasing momentum from about 60\% to 10\%, respectively.
Measurements were performed wit a setup of four radiators+Si-strip detector modules placed in two
different magnetic fields. The radiators were irregular carbon-fiber
radiators (SIGRAFIL C).

It is clear that, in order to allow for an unambiguous association
of the TR signal to the parent charged particle, this interesting concept can 
only be exploited for detectors operated in setups where the occupancy is very small.
Space-borne instruments are obvious candidates for this technology.

%\clearpage
\section{Selected modern implementations of TRDs}\label{sect:modern_trds}

In this section we present recent implementation of TRDs, with special
emphasis on two different TRD systems in large detectors at the LHC,
ATLAS and ALICE.  These two systems illustrate two complementary
approaches, dictated by their respective requirements: i) the ATLAS
TRT is a fast detector with thin detector layers realized with straw
tubes, with many layers and moderate granularity, designed for
operation in very high-rate pp collisions; ii) the ALICE TRD is a
slower detector, with thick radiators and drift chambers arranged in 6
layers, with very high granularity, optimized for Pb+Pb collisions.
In addition, we shall briefly discuss the TRDs for a fixed-target
experiments (HERMES, CBM), as well as that of the AMS experiment as an
emblematic system for a space-borne experiment.

\subsection{ATLAS TRT}\label{sect:atlas}

\subsubsection{General design}

\begin{figure}[htb]
\centering\includegraphics[angle=-90,width=0.9\textwidth]{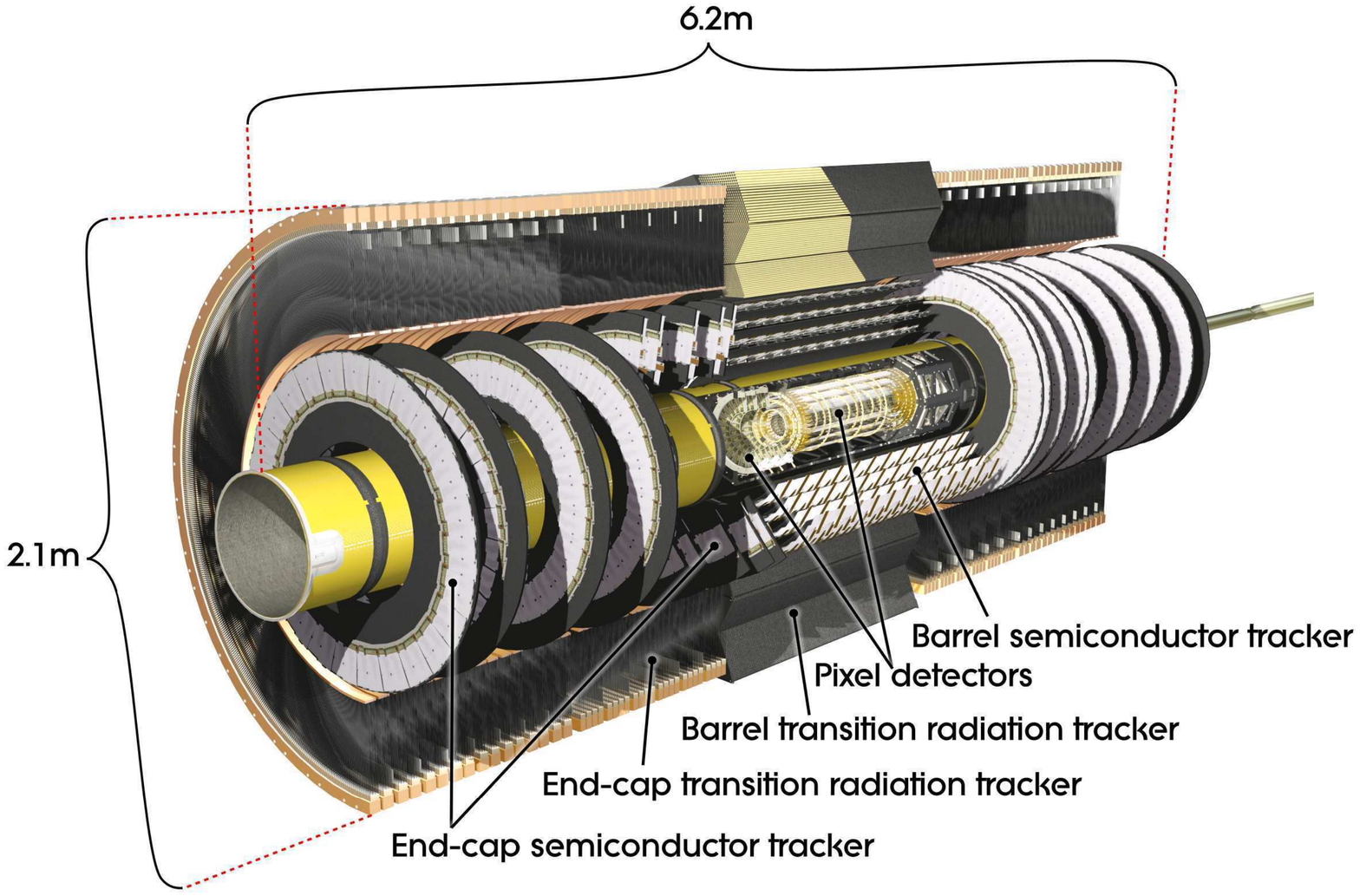} 
\caption{Schematic cutaway view of the ATLAS inner barrel
  detectors comprising the innermost pixel detector, the
  semiconductor barrel and end-cap tracker(SCT) along with the barrel and
  end-cap transition radiation tracker(TRT) (from~\cite{atlas-lhc}).}
\label{atlas-inner}
\end{figure}

The ATLAS Transition Radiation Tracker (TRT) is part of the ATLAS
central tracker, the inner detector, depicted in Fig.~\ref{atlas-inner}.
It is designed to operate in the 2~T field of
the ATLAS solenoid, where it provides both tracking information and
particle identification at the design luminosity of the LHC of
$\cal{L}$=$10^{34}\,cm^{-2}s^{-1}$. At this luminosity up to 22
overlapping events are expected in a single bunch crossing occurring
every 25~ns. 

\begin{figure}[hbt]
\begin{tabular}{lr} \begin{minipage}{.41\textwidth}
%barrel
\centering\includegraphics[width=0.95\textwidth]{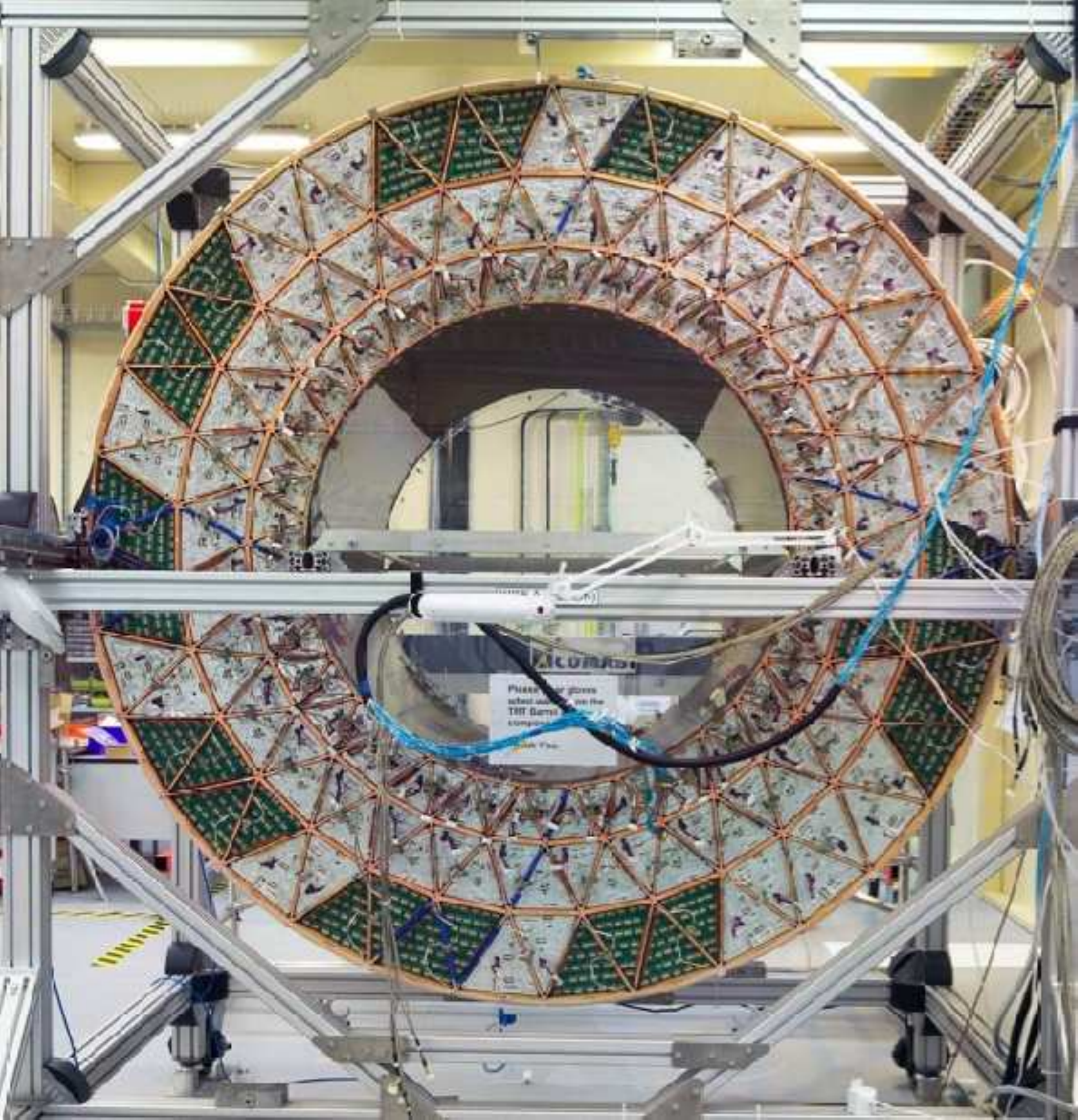}
\caption{TRT barrel during final attachment of cooling and electrical
  services (from~\cite{atlas-lhc}).}
\label{barrel-photo} 
\end{minipage} &\begin{minipage}{.55\textwidth}\vspace*{-4mm}
%endcap
\centering\includegraphics[width=0.74\textwidth,angle=-90]{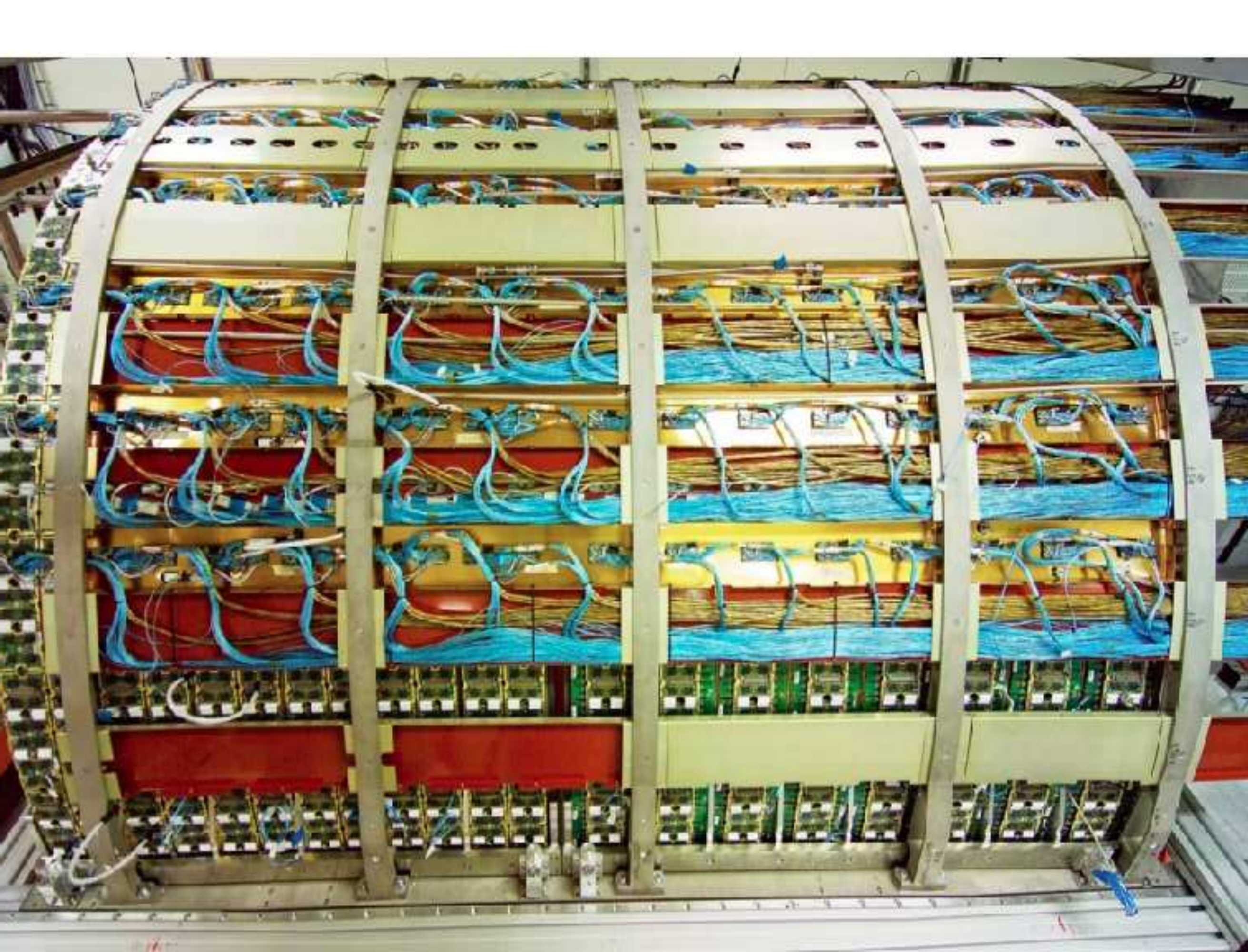}
\caption{Completed TRT end-cap during final service integration 
(from~\cite{atlas-lhc}).}
\label{endcap-photo} 
\end{minipage} \end{tabular}
\end{figure} 

The TRT itself is subdivided into two section, the TRT
barrel ($|\eta|<1.0$) and the TRT end-caps
($1.0<|\eta|<2.0$)~\cite{atlas-barrel,atlas-endcap}. The TRT barrel
has the sensor layers running parallel to the beam axis, while the
sensor layers of the end-cap TRT are radially
oriented. Fig.~\ref{barrel-photo} shows the TRT barrel, while 
Fig.~\ref{endcap-photo} shows the TRT end-cap
prior to installation. Typically, the TRT provides 36 hits per track
with a precision of about 140~$\mu$m in the bending direction. The
combination of the precision inner tracker and the hits in the TRT
contribute to the precision momentum measurement and robust pattern
recognition of the ATLAS detector.

\begin{table}[htb]
\begin{center}
\caption{Synopsis of the ATLAS TRT parameters}
\label{atlas:tab:overview} 
\begin{tabular}{|l|l|l|}
\hline
 & TRT barrel & TRT end-cap \\\hline
 Pseudo-rapidity coverage & $|\eta|<1.0$ & $1.0<|\eta|<2.0$ \\
 Position along beam axis & $|z|<712$~mm & $848<|z|<2710$~mm  \\
 Radial position & $563<r<1066$~mm & $644<r<1004$~mm  \\
 Total weight & 707~kg & 2$\times$1120~kg\\
\hline 
 Number of straw planes & 73 & 160 \\
 Number of layers & 73 & 160 \\
 Length of straws & 144~cm & 37~cm \\
 Total number of straws& 52544 & 122880 \\
 Radiator & fibers & 15~$\mu$m PP %polypropylene 
foils with spacer \\
 Radiation length & $0.2\,X/X_0$ & $\approx 0.6\,X/X_0$ \\
\hline 
 Gas volume & $\simeq$1 m$^3$ & 0.6 m$^3$ \\\hline
 Detector gas & \multicolumn{2}{c|}{Xe/CO$_2$/O$_2$ (70\%/27\%/3\%)} \\
 Straw diameter &\multicolumn{2}{c|}{4~mm}\\
 Gas gain & \multicolumn{2}{c|}{$2.5\cdot 10^4$}\\
 Cathode voltage & \multicolumn{2}{c|}{-1530~V}\\
 Drift velocity &  \multicolumn{2}{c|}{$52$~$\mu$m/ns}\\
\hline 
 Number of readout channels & 105088 & 245760 \\
\hline
 Counting rate per wire & \multicolumn{2}{c|}{up to 20~MHz}\\
 Average number of hits per wire & \multicolumn{2}{c|}{22-36}\\
 Average number of TR hits & \multicolumn{2}{c|}{5-10 (for electrons), 2
   (for $\pi$)}\\
\hline 
\end{tabular}
\end{center}
\end{table}

\subsubsection{Detector layout}

The ATLAS TRT is based on straws, which in case of the barrel are
144~cm long. They are electrically separated into two halves at
$\eta=0$ and arranged in a total of 73 planes. The end-cap straws are
37~cm long, radially arranged in wheels with a total of 160 planes.
The straws themselves are polyimide tubes with a diameter of 4~mm.
Its wall is made of two 35~$\mu$m thick
multi-layer films bonded back-to-back. The film comprises a 25~$\mu$m
thick polyimide film with a 200~nm Al layer protected by a 5-6~$\mu$m 
thick graphite-polyimide layer. The backside of the film is coated 
with a 5~$\mu$m polyurethane layer used to
heat-seal the two films back-to-back.
Carbon fibers along the straws ensure their stability. The anodes
are 31~$\mu$m diameter gold-plated tungsten wires.
They are directly connected to the front-end electronics and kept at
ground potential. The anode resistance is approximately 60~$\Omega$/m and the
assembled straw capacitance is $<$10 pF. The signal attenuation length
is $\approx$~4 m and the signal propagation time is $\approx$~4 ns/m. 
The straws are operated at a gain of $2.5\cdot 10^4$ with a
gas mixture of Xe/CO$_2$/O$_2$(70:27:3) and a slight overpressure with
respect to atmospheric pressure of 5-10~mbar. 
Under normal operating conditions, the maximum electron
collection time is $\approx$48~ns and the drift-time accuracy leads to
a position resolution in bending direction of about
130~$\mu$m\cite{atlas-straws}. TR photons are
absorbed in the Xe-based gas mixture, and yield much larger signal
amplitudes than minimum-ionizing charged particles. The distinction
between TR and tracking signals is obtained on a straw-by-straw basis
using separate low and high thresholds in the front-end electronics.
For the barrel straws, the anode wires 
are read out on both ends. Close to center, the
wires are supported mechanically by a plastic insert glued to the
inner wall of the straw. The wires are electrically separated by a fused glass
capillary of 6~mm length and 0.254~mm diameter to cope with the
occupancy. This leads to a local inefficiency of about 2~cm around the
center of the straw. For the innermost nine layers of the wires are
subdivided into three segments, which renders only the 31.2 cm-long 
end-segments on each side active.  

\begin{figure}
\centering
\includegraphics[angle=-90,width=0.8\textwidth]{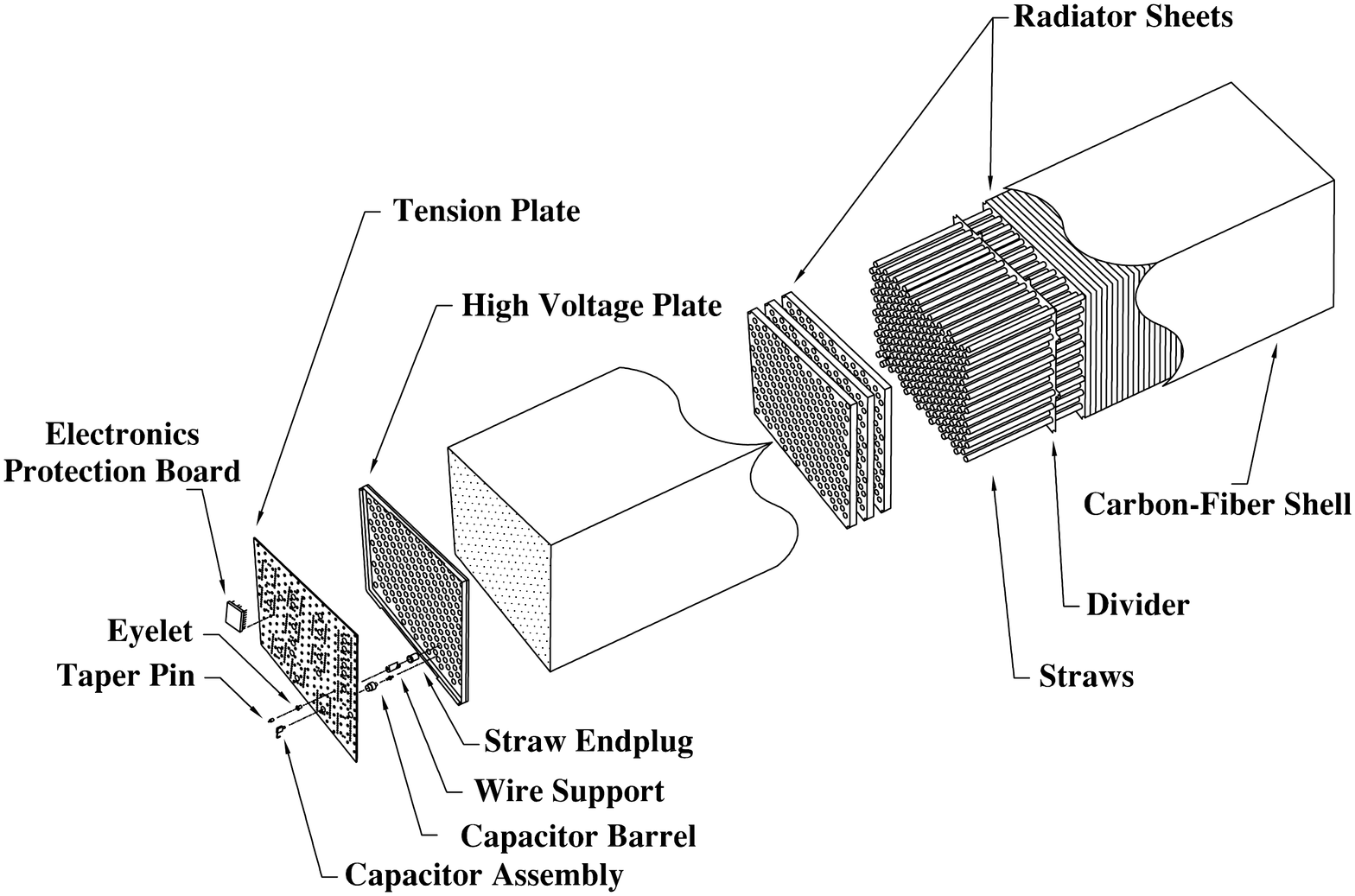}
\caption{Isometric view of a ATLAS TRT barrel module (from~\cite{atlas-barrel}).}
\label{barrel-explosion}
\end{figure}

For stable operation, the wire offset with respect
to the straw center is needs to be $<300\,\mu$m. Since the wire sag
itself is $<15\,\mu$m this
translates directly into the requirement on the straightness of the straws.
To maintain straw straightness in the barrel, alignment planes made of 
polyimide with a matrix of holes are positioned each 25~cm along the 
z-direction of the module. A schematic drawing of the assembly of TRT barrel 
module is shown in Fig.~\ref{barrel-explosion}.
Stable operation of the ATLAS TRT straws with the Xe-based gas mixture 
requires a re-circulating gas system with continuous monitoring of the 
gas quality. To avoid pollution from permeation through the straw walls 
or through leaks, the straws are operated within an envelope of CO$_2$.

\subsubsection{Electronics}

\begin{figure}[htb]
\centering
\includegraphics[angle=-90,width=0.9\textwidth]{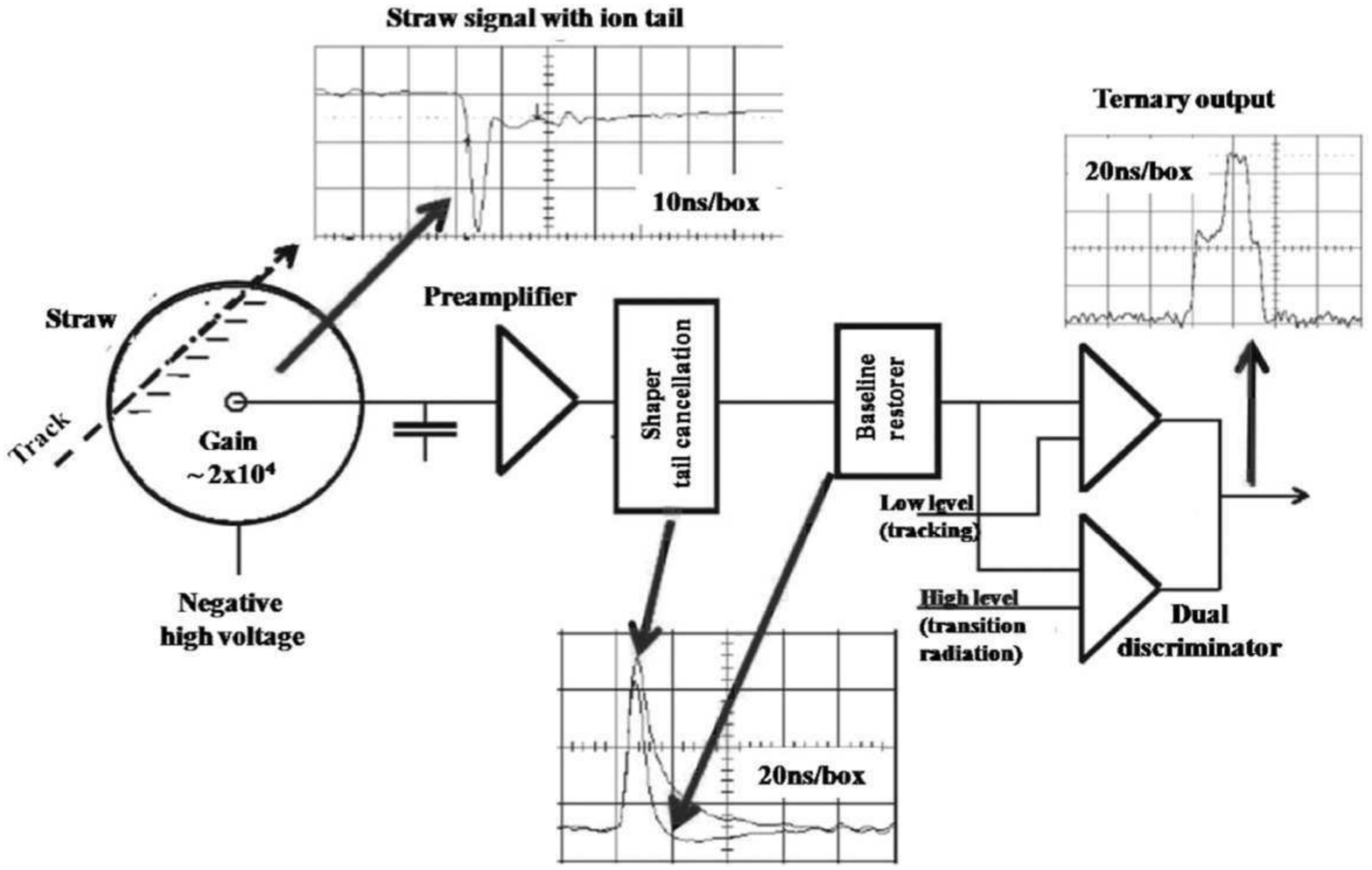} 
\caption{Schematic drawing of the ATLAS TRT electronics readout chain,
  along with the detector signal after amplification and shaping, the
  baseline restoration and the dual discriminator (from~\cite{atlas-lhc}).}
\label{trt-electronics}
\end{figure}

The ATLAS TRT electronics readout chain along with typical signals at the various
stages of signal processing are shown in Fig.~\ref{trt-electronics}.
The analog signal processing and threshold discrimination to detect
signals from both minimum-ionizing particles for tracking and transition
radiation from electrons 
as well as the subsequent time digitization and data pipelining 
are implemented in two ASICs, which are directly mounted on the
detector. 
Each stage of the TRT signal readout chain
comprises: (1) an eight-channel analog ASIC~\cite{atlas-pasa}, called the
ASDBLR realized in bi-CMOS radiation tolerant DMILL technology. It
accounts for amplification, shaping, and baseline restoration. It
includes two discriminators, one operating at low threshold (typically
250 eV) for signals from minimum-ionizing particles and one operating at
high threshold (typically 6 keV) for transition radiation detection.
(2) a 16-channel ASIC fabricated using commercial
radiation tolerant 0.25~$\mu$m CMOS technology~\cite{atlas-dtmroc}. 
This ASIC performs the drift-time measurement ($\approx$3~ns
binning). 
It includes a digital
pipeline for holding the data during the L1 trigger latency, a
derandomising buffer and a 40~Mbits/s serial interface. It also
includes the necessary interface to the timing, trigger and control as
well as DACs to set the discriminator thresholds of the analog ASIC
along with further test circuitry. These ASICs are mounted on
front-end boards directly attached to the detector. The electronics 
is cooled by a liquid mono-phase fluorinated (C$_6$F$_{14}$) cooling system.

At the TRT operating low threshold used for tracking (equivalent to
$\approx$15\% of the average signal expected from minimum-ionizing particles),
the mean straw noise occupancy is about 2\%.
The expected maximum straw occupancy is 50\%. The entire front-end electronics 
chain was exposed to a neutron dose of about $4\cdot 10^{14}$~cm$^{-2}$ 
and to a $\gamma$-ray dose of 80~kGy. 
Variations of up to 25\% were observed in the ASDBLR gain, however, with no change 
in the effective thresholds and noise performance after a standard voltage
compensation procedure.

\subsubsection{ATLAS TRT performance}

As mentioned above the TRT plays a central role within ATLAS for
electron identification, cross-checking and complementing the
calorimeter, especially at momenta below 25~GeV/c. In addition, the TRT
contributes to the reconstruction and identification of electron
track segments from photon conversions down to 1~GeV and of electrons
which have radiated a large fraction of their energy in the silicon
layers.

\begin{figure}[htb]
\centering
\includegraphics[angle=-90,width=0.9\textwidth]{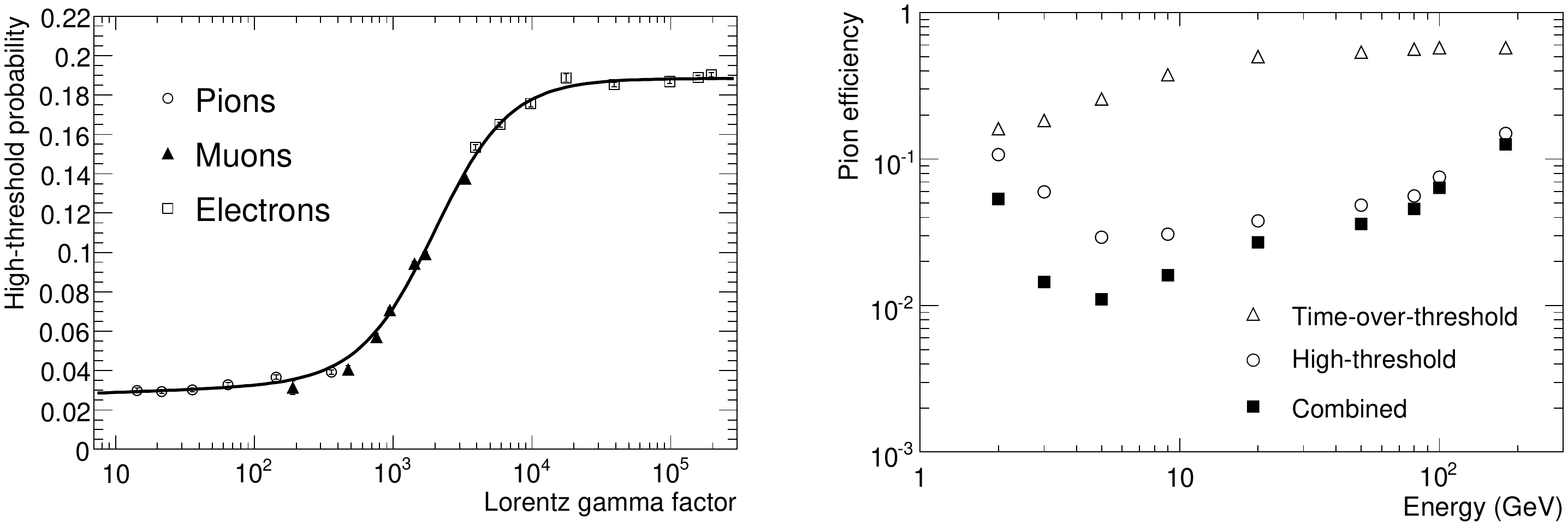} 
\vspace{-2cm}
\caption{Left panel) Average probability of a high-threshold hit in
  the barrel TRT as a function of the Lorentz $\gamma$-factor for
  different particle species. Right panel) Pion efficiency (determined
  at 90\% electron efficiency) as a
  function of pion energy using different discrimination techniques
  (cf. text)(from~\cite{atlas-lhc}).}
\label{trt-gamma-effi}
\end{figure}

In a combined testbeam the response of the ATLAS inner barrel to pions,
electrons and muons in the momentum range between 2
and 350 GeV/c has been evaluated. In Fig.~\ref{trt-gamma-effi}
the high-threshold hit probability towards different particle species
is shown (left panel). TR contributes significantly to the
high-threshold hits for electron momenta above 2~GeV/c and saturation
sets in for electron momenta above about 10~GeV/c. The right panel of
that figure shows the
resulting pion identification efficiency for an electron efficiency of
90\%, achieved by performing a likelihood evaluation based on the
high-threshold probability for electrons and pions for each
straw. It also demonstrates how the inclusion of
time-over-threshold information (which quantifies the energy deposit 
in the straw) improves the pion rejection by about a factor of two 
when combined with the high-threshold hit information. 
The pion rejection power reaches a maximum at momenta of $\approx$5~GeV/c. 
In general, pion rejection factors above 50 are achieved in the energy range 
of 2-20 GeV. 

\begin{figure}[htb]
\centering
\includegraphics[angle=-90,width=0.6\textwidth]{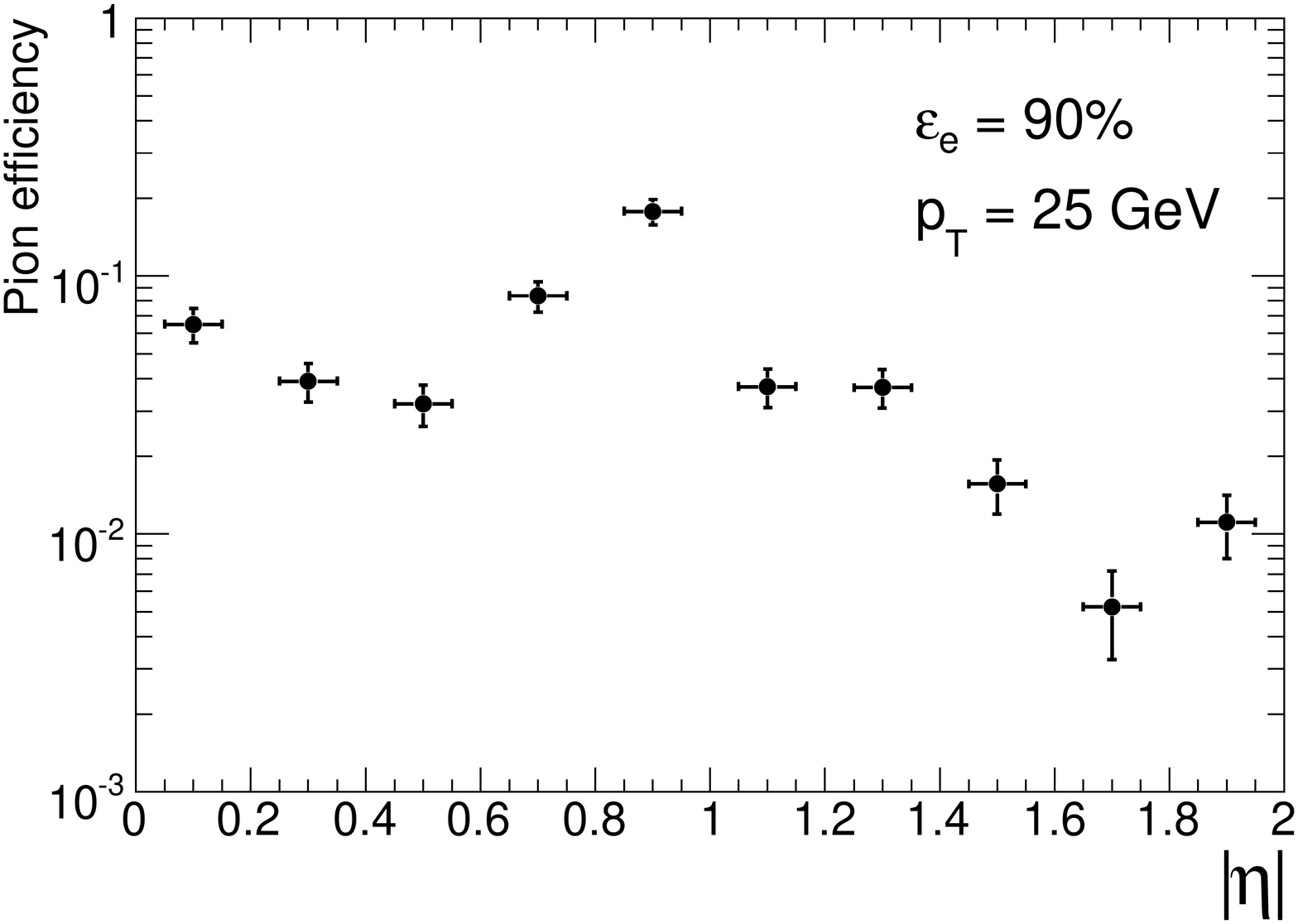} 
\caption{Expected pion efficiency as a function of pseudo-rapidity at
  90\% electron efficiency for electrons with a transverse momentum
  of $p_t=$25~GeV/c (from~\cite{atlas-lhc}).}
\label{trt-pion-effi}
\end{figure}

The electron-pion separation expected for the ATLAS TRT, including the
aforementioned time-over-threshold information, is shown as a function of
pseudo-rapidity in Fig.~\ref{trt-pion-effi} as the pion identification 
efficiency expected for an electron efficiency of 90\%. The shape observed 
is strongly correlated with the number of TRT straws crossed by the track.
It decreases from approximately 35 to a minimum of 20 in the
transition region between the barrel and end-cap TRT,
$0.8<|\eta|<1.1$. 
It also decreases rapidly at the edge of the
TRT fiducial acceptance, which is limited to $|\eta|>1.8$. 
Since the TR yield depends on momentum and these results are for fixed
transverse momentum $p_t$,
part of the $\eta$ dependence arises from the momentum dependence 
of the TR yield.
Owing to its more efficient regular foil radiator, the performance in 
terms of particle identification is better in the end-cap TRT than in 
the barrel TRT~\cite{atlas-lhc}.

\subsection{ALICE TRD}\label{sect:alice}

\subsubsection{General design}
\label{alice:design}

The purpose of the ALICE TRD~\cite{ali} is twofold. On the one hand, it 
provides efficient electron identification in the central barrel for momenta
above 1 GeV/c. On the other hand, based on its inherent
tracking capability, the readout is able to provide a fast
trigger for charged particles with high momenta. In
conjunction with data from the Inner Tracking System (ITS) and the
Time Projection Chamber (TPC) it is possible to study
the production of light and heavy vector-mesons and the
dilepton continuum both in p-p as well as in Pb-Pb collisions.
The trigger will be used for jet studies and to
significantly enhance the recorded $\Upsilon$-yields,
high-p$\,_t$ $J/\psi$, and the high-mass part of the dilepton continuum.

\begin{figure}[htb]
\centering\includegraphics[width=.7\textwidth]{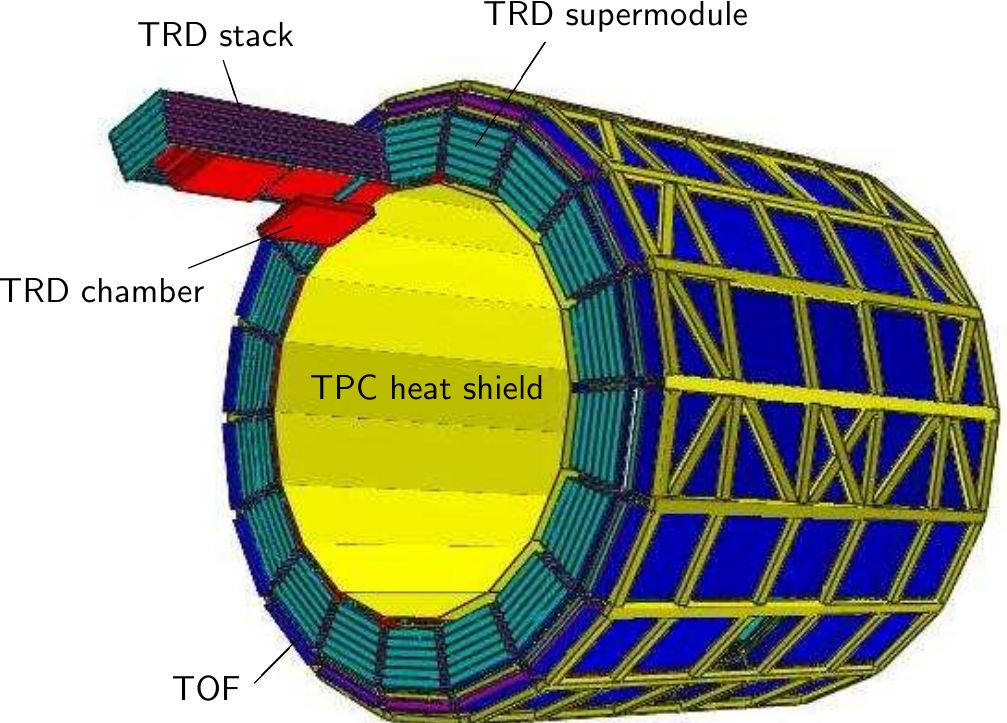}
\caption{Schematic drawing of the TRD layout in the ALICE space
  frame (courtesy D. Mi\` skowiec). 
  Shown are 18 super modules (light blue face side) each
  containing 30 readout chambers (red) arranged in five stacks of six
  layers. One chamber has been displaced for clarity. On the outside
  the TRD is surrounded by the time-of-flight (TOF) system (dark
  blue). On the inside the heat shield (yellow) towards the TPC is
  shown. In radial direction the TRD occupies the range 2.9-3.7 m and 
  the length is almost 8 m.} 
\label{alice:TRDlayout}
\end{figure}

The design parameters of the TRD were driven by the physics
considerations~\cite{alice-trdtdr}:

\begin{itemize}
\item []{\em Pion rejection capability} - this is governed by the
  signal-to-background ratio in the measurement of $J/\psi$ production
  and its p$_t$ dependence. This led to the design goal for the pion
  rejection capability of a factor 100 for momenta above 1
  GeV/c \cite{alice-trdtdr}, which is necessary for the
  measurement of the lighter vector-mesons and the determination of
  the continuum between the $J/\psi$ and the $\Upsilon$. 

\item []{\em Position and momentum resolution} - needs to be of the
  order of a fraction of a TPC pad to match and exploit the combined
  momentum resolution leading to an overall mass resolution of about
  100 MeV/c$^2$ at the $\Upsilon$-mass. The
  anticipated momentum resolution of the TRD itself at 5
  GeV/c of 3.5\% (4.7\%) for low (high) multiplicity will crucially 
determine the sharpness of the trigger threshold in p$\,_t$ as well as the
  capability to reject fake tracks.

\item []{\em Radiation length} - has to be minimized in order to
  reduce Bremsstrahlung leading to incorrect momentum determination or
  loss of electrons and to reduce photon conversions resulting in
  increased occupancy as well as incorrect matching.

\item[] {\em Detector granularity} - in bending direction it is
  governed by the desired momentum resolution and in longitudinal
  direction by the need to correctly identify and track electrons
  through all layers of the detector even at the largest anticipated
  multiplicities. This led to pads with an average area of about 6
  cm$^2$. With this a tracking efficiency of 90\% can be achieved for
  single tracks at a maximum occupancy of 34\% including secondaries
  at the highest simulated multiplicity density of $\rm d
  N_{\textrm{ch}}/\rm d\eta$ = 8000.
\end{itemize}

%----------------------------------------------------------
\subsubsection{Detector layout}
\label{trd:Layout}

\begin{figure}[h]
\centering\includegraphics[width=.55\textwidth]{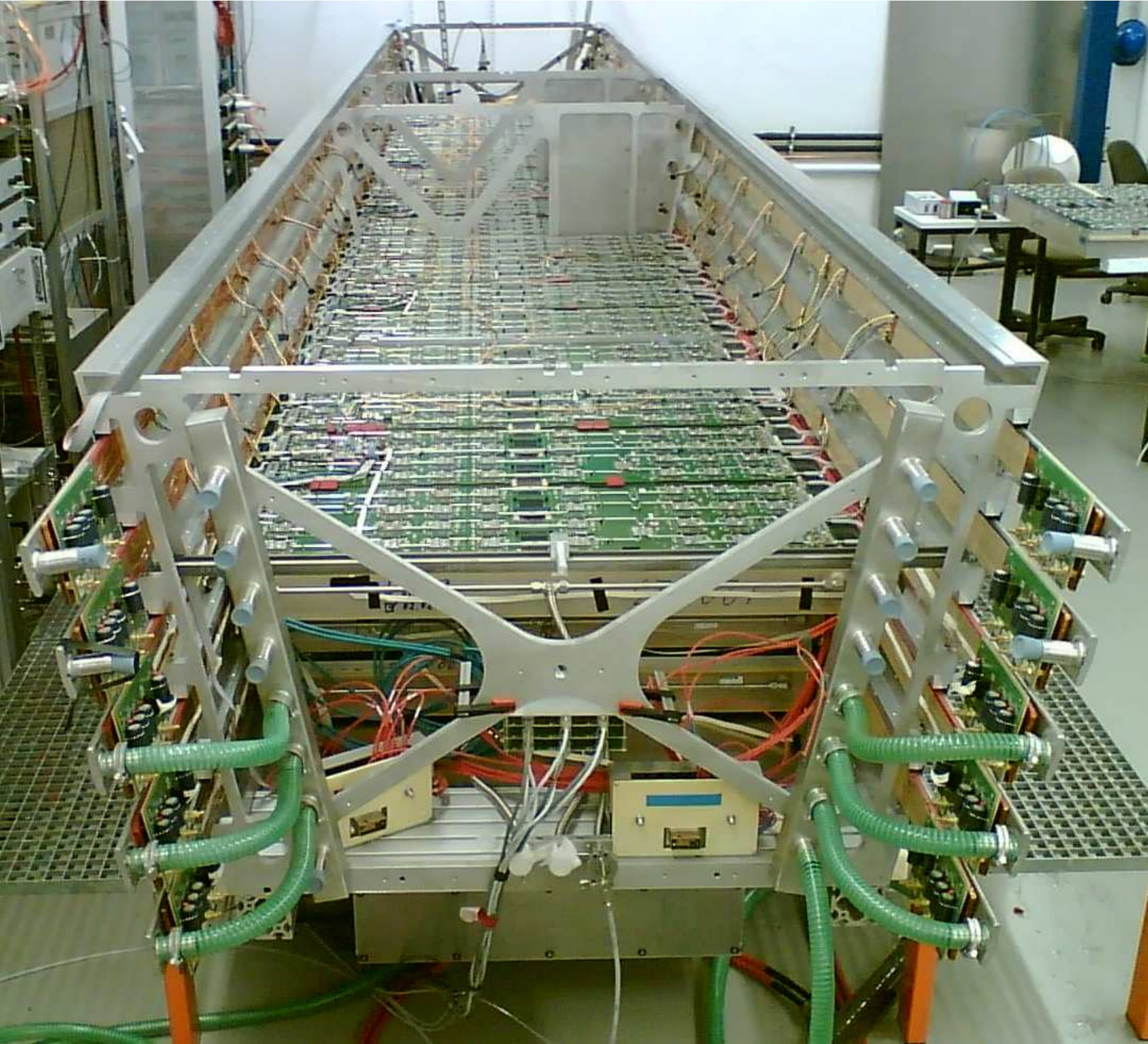}
\caption{Super module during assembly with the chambers of the first three 
layers installed.} 
\label{alice:TRDsm}
\end{figure}

  The final design of the TRD is depicted in
  Fig.~\ref{alice:TRDlayout}. The TRD consists of 540 individual readout
  detector modules. They are arranged into 18 so called super modules
  (Fig.~\ref{alice:TRDsm})
  each containing 30 modules arranged in five stacks and six
  layers. In longitudinal ($z$) direction the active length is 7~m,
  the overall length of the entire super module (Fig.~\ref{alice:TRDsm}) 
  is 7.8~m, its weight is about 1700~kg. 
  
  Each detector element consists of a carbon fiber laminated
  Rohacell/polypropylene fiber sandwich radiator of 48~mm thickness, a
  drift section of 30~mm thickness, and a multi-wire proportional
  chamber section (7~mm) with pad readout. The pad planes are
  supported by a honeycomb carbon-fiber sandwich back panel
  (22~mm). While very light, the panel and the radiator provide enough
  mechanical rigidity of the chamber to cope with overpressure up to 
  1~mbar to ensure a deformation of less than 1~mm.  The entire readout
  electronics is directly mounted on the back panel of the
  detector. Including the water cooling system the total thickness of
  a single detector layer is 125~mm. In the bending plane ($r\varphi$)
  each pad row consists of 144 pads. The central chambers consist of
  12, all others of 16 pad rows. This leads to an overall channel
  count of 1.18$\times$10$^6$. The total active area subtended by the
  pads is 716~m$^2$. 
The construction parameters, operating conditions and design performance
of the detector are summarized in Table~\ref{alice:tab:overview}.
  
\begin{table}[h!]
\begin{center}
\caption{Synopsis of the ALICE TRD parameters}
\label{alice:tab:overview} 
\begin{tabular}{|l|l|}
\hline
 Pseudo-rapidity coverage & $-0.84<\eta<0.84$ \\
 Radial position & $2.90<r<3.68$~m \\\hline 
 Largest module & $117\times 147$~cm$^{2}$ \\
 Active detector area & 716 m$^{2}$ \\
 Radiator & fiber/foam sandwich, 4.8 cm per layer \\
 Radial detector thickness (for $|z|>$ 50~cm) & $X/X_{0}=21.6\%(25.7\%)$ \\
\hline 
 Module segmentation & 144 in $\varphi$ \hskip10mm 12--16 in $z$ \\
 Typical pad size & $0.7\times8.8$~cm$^{2}$ \\
 Total number of pads & $1.18\times10^{6}$ \\
\hline 
 Detector gas & Xe/CO$_2$ (85\%/15\%) \\
 Gas volume & $27.2$~m$^{3}$ \\
 Depth of drift region & $3$~cm \\
 Depth of amplification region & 0.7~cm \\
 Drift field &  $0.7$~kV/cm \\
 Drift velocity &  $1.5$~cm/$\mu$s\\
\hline 
 Number of readout channels & $1.18\times10^{6}$ \\
 Time samples in $r$ (drift) & 24 \\
 ADC & 10 bit, 10 MHz\\
 Number of multi-chip modules & 70848\\
 Number of readout boards & 4104 \\
\hline
 Pad occupancy for $\mathrm{d}N_{ch}/\mathrm{d}\eta=8000$ &  $34\%$ \\ 
 Pad occupancy in pp & $2\times 10^{-4}$ \\
 Space-point resolution at $1$~GeV$\,c^{-1}$ \hfill in $r\varphi$ & $400 (600)$~$\mu$m for $\mathrm{d}N_{ch}/\mathrm{d}\eta=2000$ (8000)\\
 \hfill in $z$\ \ \  & $2$~mm (offline) \\
 Momentum resolution & $\delta p/p = 2.5\% \oplus 0.5\% (0.8\%)p$/(GeV/c) \\
& for $\mathrm{d}N_{ch}/\mathrm{d}\eta=2000$ (8000) \\ 
 Pion suppression at $90\%$ electron efficiency & better than 100 for $p \geq$ 1~GeV/c\\
\hline 
 Event size for $\mathrm{d}N_{ch}/\mathrm{d}\eta=8000$ & 11 MB \\
 Event size for pp & 6 kB \\
% Trigger rate limit & 100 kHz \\
  Rate limitit for triggering & 100 kHz \\
\hline
\end{tabular}
\end{center}
\end{table}

\begin{figure}[htb] 
\begin{tabular}{cc} \begin{minipage}{.48\textwidth}
\centering\includegraphics[width=.9\textwidth]{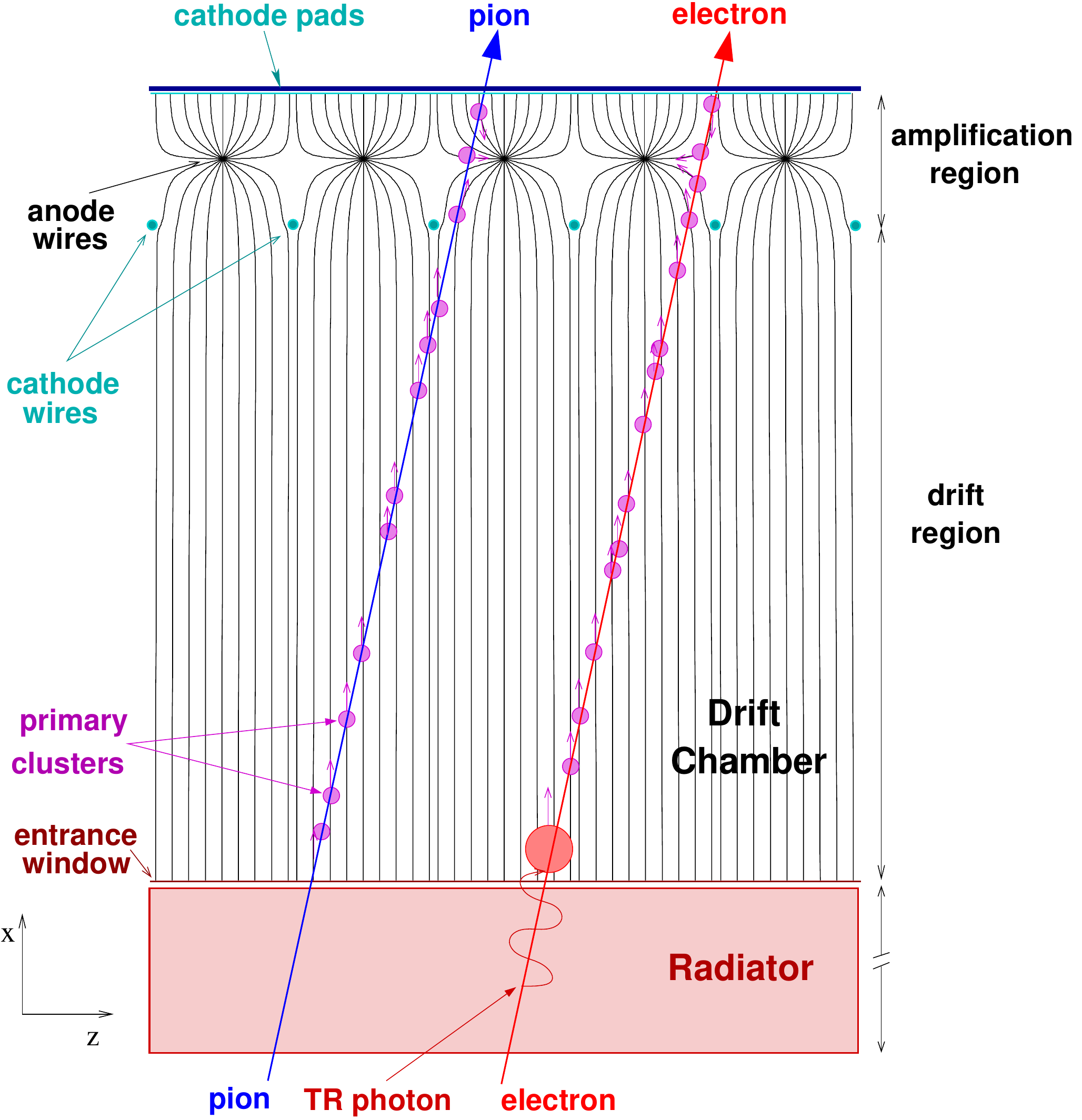}
\end{minipage} &\begin{minipage}{.48\textwidth}
\centering\includegraphics[width=.9\textwidth]{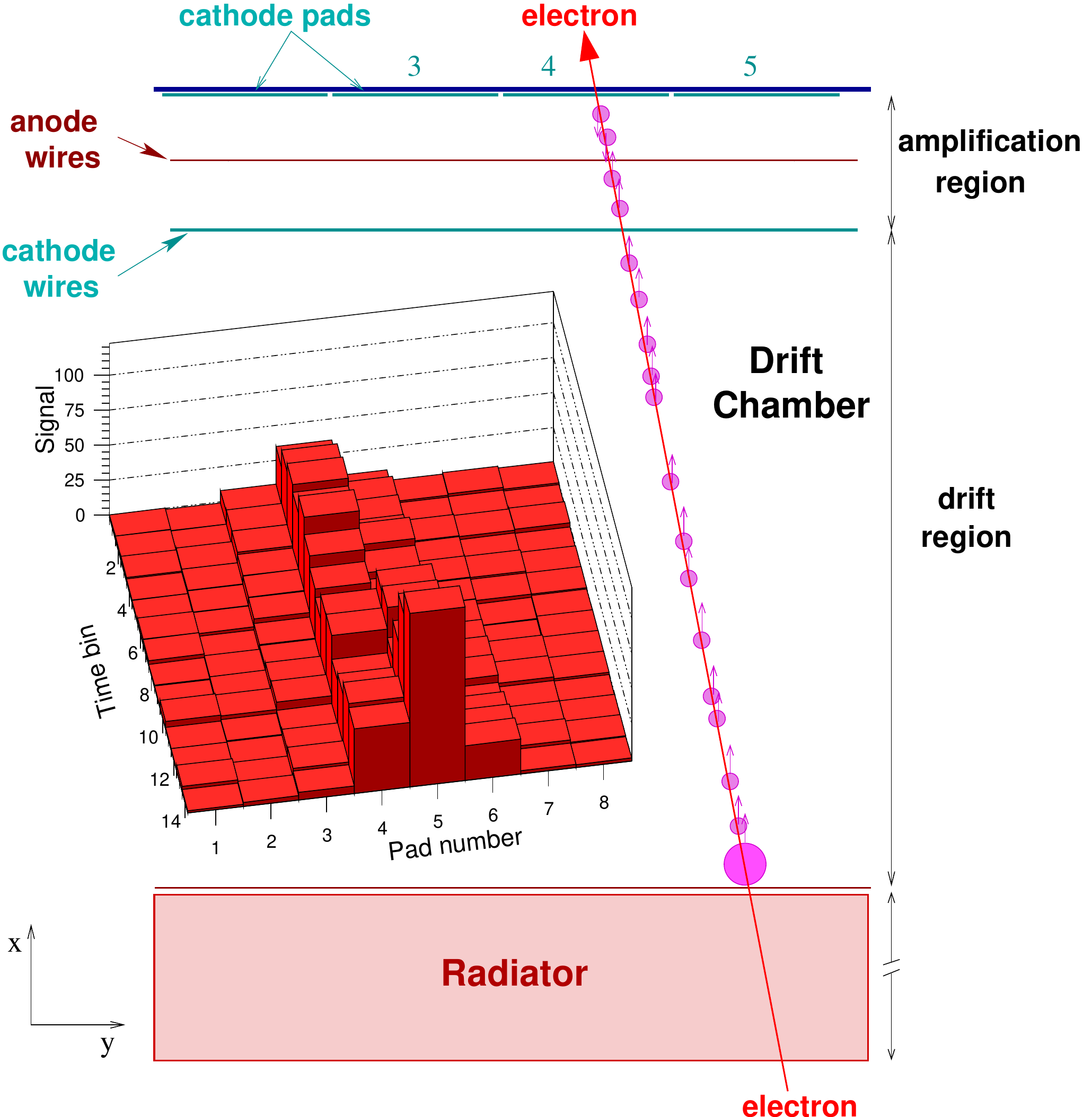}
\end{minipage} \end{tabular}

\caption{Schematic cross-sectional view of a detector module in
  $rz$-direction (left panel) and $r\varphi$-direction (right
  panel). The inset shows the charge deposit from an inclined track.}
\label{alice:TRDprinciple}
\end{figure}

Cross-sectional views of a single TRD chamber are shown in 
Fig.~\ref{alice:TRDprinciple}. Ionizing radiation produces electrons 
in the counting gas (Xe/CO$_2$ (85:15)). 
Particles exceeding $\gamma \approx 1000$ will in addition
produce about 1.45 X-ray photons in the energy range of 1 to 30
keV. The largest conversion probability for TR is at the
very beginning of the drift region. All electrons from ionization
energy loss and X-ray conversions will drift towards the anode
wires. Following gas amplification the signal is induced on the readout
pads. A typical track is shown in the inset of the central panel of
Fig.~\ref{alice:TRDprinciple}.  The inclination of the track in
bending direction is a direct measure of its transverse momentum.  

\begin{figure}[h] 
\centering\includegraphics[width=.53\textwidth,height=.5\textwidth]{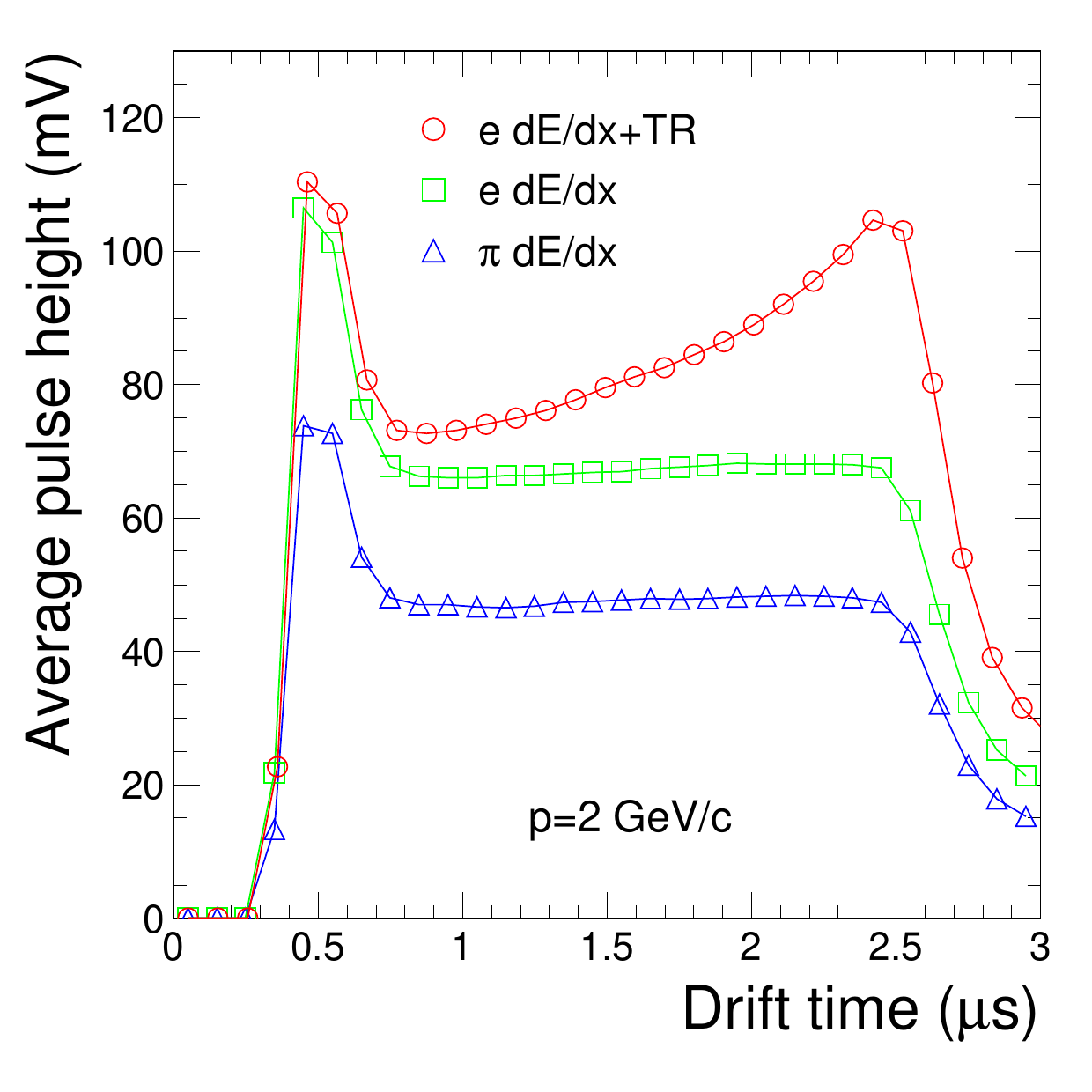}
\caption{The  average pulse height as a function of drift time for pions
  (triangles), electrons without a radiator (squares) and electrons
  with a radiator (circles) for 2 GeV/c momentum.}
\label{alice:TRDsig}
\end{figure}

For particles with a momentum of 2 GeV/c the average amplitude of
the cathode pad signal versus drift time is shown in  Fig.~\ref{alice:TRDsig}. 
The conversion of the TR right at the entrance of the chamber,
i.e. at large drift times, is clearly visible.

\subsubsection{Readout electronics}

\begin{figure}[htb]
\centering
\includegraphics[width=\textwidth]{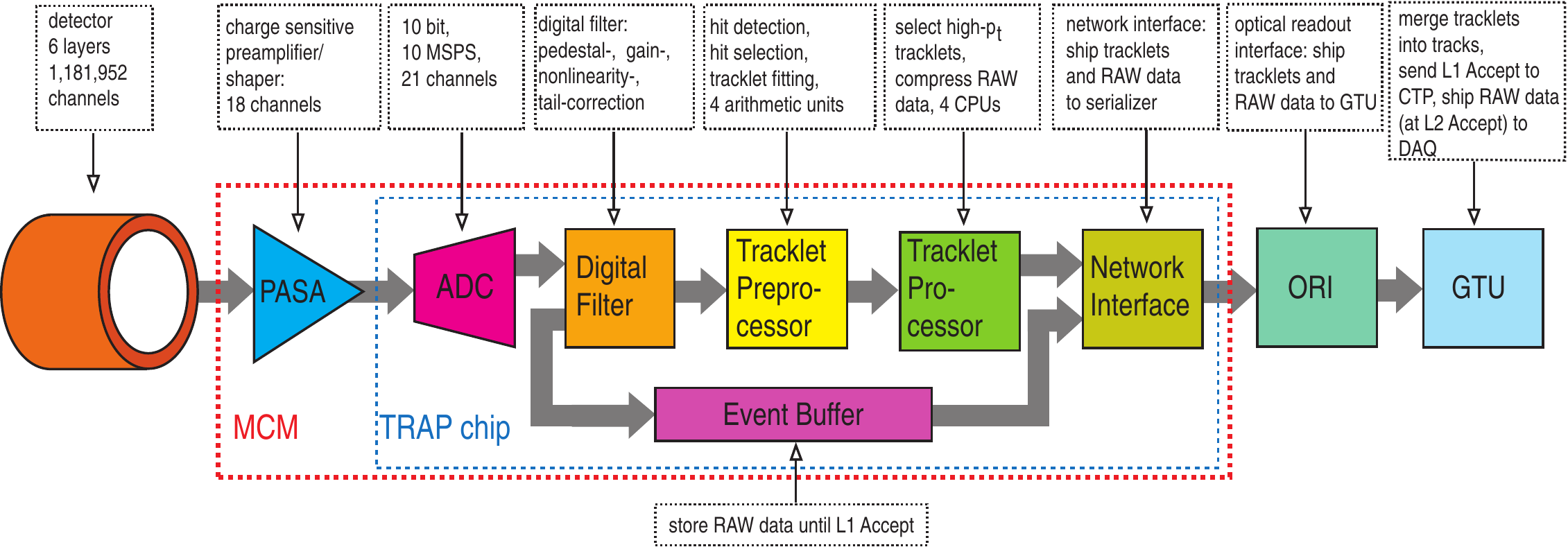} 
\caption{Schematic overview of the readout electronics of the TRD.}
\label{alice:electronics}
\end{figure}

An overview of the ALICE TRD readout electronics \cite{alice-fee} is shown 
in Fig.~\ref{alice:electronics}. The electronics including the optical
serializers ORI (two per chamber) is directly mounted on the backside
of the detector modules. The data are transmitted to the Global
Tracking Unit (GTU) via 60 optical fibers per super module.  The GTU
either passes the data directly to the DAQ 
or further processes the data in order to derive a fast Level
1 trigger decision. In that case individual tracklets from different
layers of a stack are combined to determine the multiplicity of
high-p$\,_t$ particles or to detect high-momentum e$^+$e$^-$-pairs. At
Level 1, after about 6.1~$\mu$s this trigger is transmitted to the 
Central Trigger Processor CTP.

Most of the on-detector readout electronics are realized as ASICs. Two
such chips, an 18-channel preamplifier shaper (PASA), and
a so-called Tracklet Processor (TRAP) have been integrated into a
multi-chip module (MCM). The PASA \cite{alice-pasa} is a folded
cascode with differential output (120 ns shaping time, 850e ENC for an
input capacitance of 25 pF, 12.4 mV/fC gain, and 12 mW/channel power
consumption). It has been realized using the AMS 0.35~$\mu$m CMOS
process. Each PASA has 18 input and 21 output channels.  For
reconstruction of tracklets the extra output channels are fed into the
analog inputs of the TRAPs on neighboring MCMs to allow for continuous
charge sharing across MCM boundaries.

The Tracklet Processor~\cite{alice-trap} is a mixed signal design 
(UMC 0.18~$\mu$m). It comprises 21 ADCs, digital filters, event
buffers, and processing units that allow to
calculate the inclination of track segments in bending direction as well as
the total charge deposited along the track (Local Tracking Unit -
LTU). This feature allows to identify high-p$\,_t$ particles on the
trigger level. Evaluation of the deposited energy will furthermore
allow to tag possible electron candidates on the trigger level.

\begin{figure}[htb]
\centering
\includegraphics[width=0.83\textwidth]{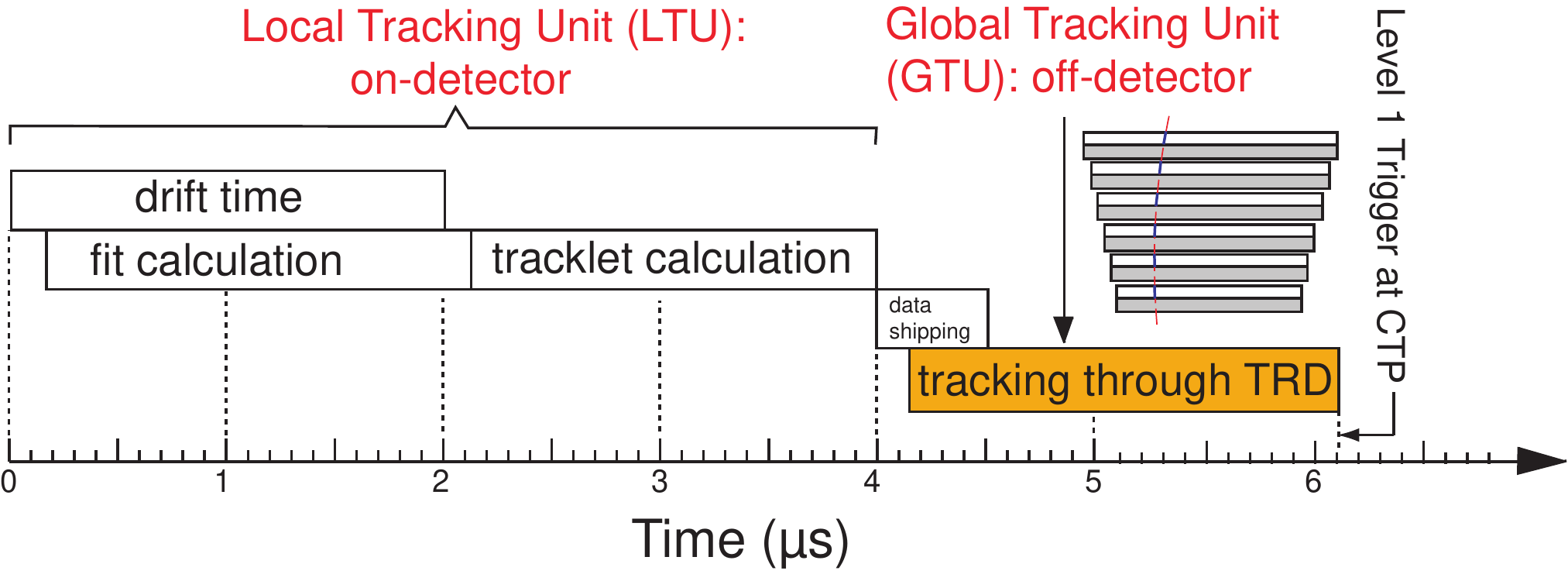} 
\caption{Trigger timing diagram for the generation of a high-p$_t$
  trigger.}
\label{alice:trigger}
\end{figure}

The resulting track segments from
the different detector layers have to be matched in three dimensions
for transverse momentum reconstruction.
Based on the data of all 1.2 million analog channels, the  
reconstruction has to be performed within 6.1 $\mu$s to derive the  
Level-1 trigger decision. The entire trigger timing sequence involving
the LTU and the GTU is depicted in Fig.~\ref{alice:trigger}. 

The massively parallel hardware architecture of the GTU is capable of
processing up to 20 000 track segments within 2~$\mu$s.
The core of the GTU, along with a custom bus system,
is the so-called Track Matching Unit (TMU).
It is an FPGA-based system utilizing PCI and 12 fiber-optical transceiver 
interfaces gathering the data from a stack of six chambers. It is 
realized as a CompactPCI plug-in card. The main FPGA is a Xilinx  
Virtex-4 FX chip which includes an integrated multi-gigabit serializer/ 
deserializer and PowerPC processor blocks.

%----------------------------------------------------------
\subsection{ALICE TRD performance}
\label{alice:performance}

\subsubsection{Specific energy loss and TR}

\begin{figure}[hbt]
\begin{tabular}{lr} \begin{minipage}{.48\textwidth}
%dedx
\vspace{-1.cm}
\centering\includegraphics[width=1.1\textwidth]{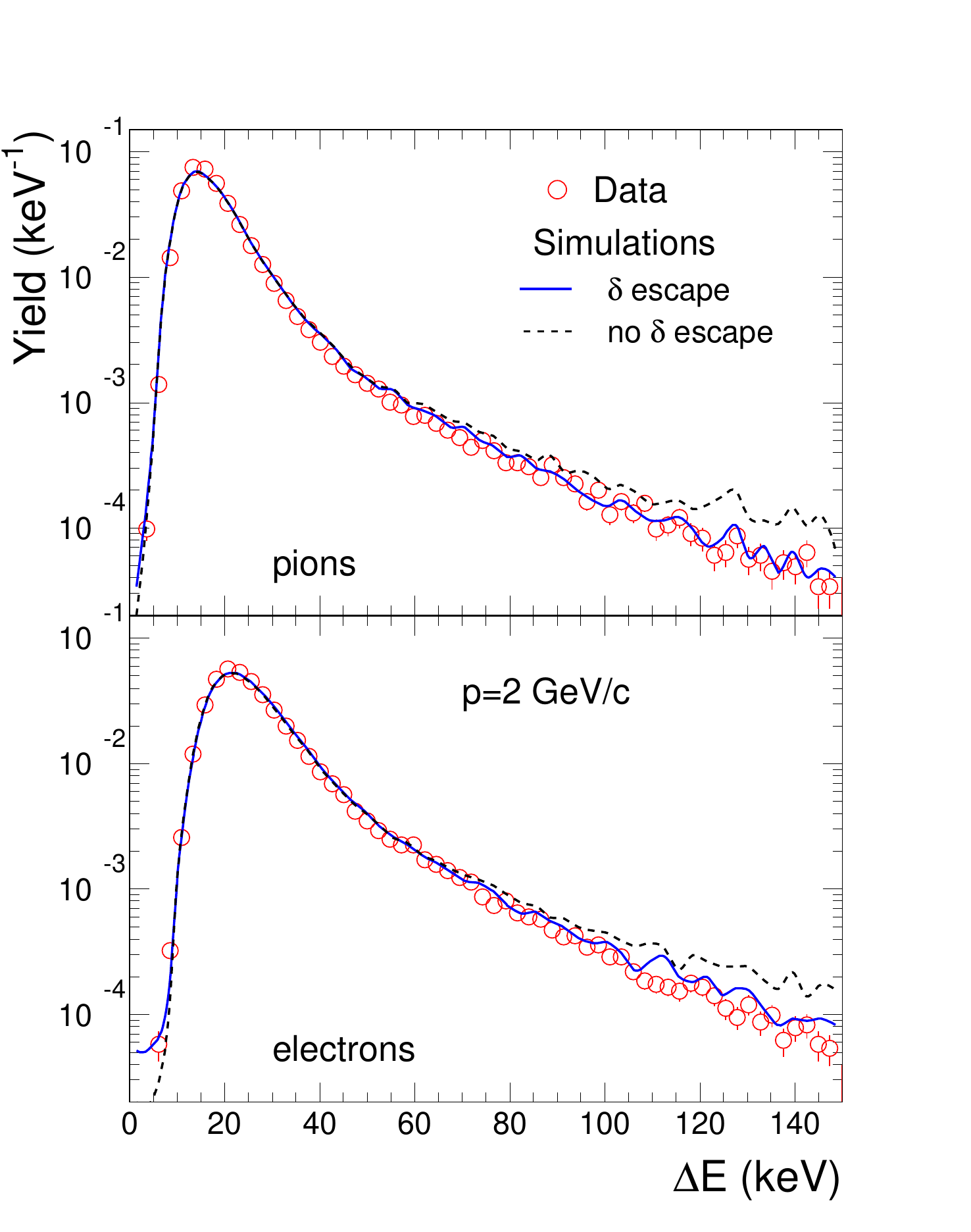}
\caption{Measurement of the specific energy deposit of 2 GeV/c 
pions (top) and electrons (bottom) in Xe,CO$_2$ (85:15) along with
simulations \cite{alice-dedx}. ``$\delta$ escape'' denotes the realistic 
treatment of $\delta$-rays.}
\label{alice:dedx} 
\end{minipage} &\begin{minipage}{.48\textwidth}
%tr
\centering\includegraphics[width=1.0\textwidth]{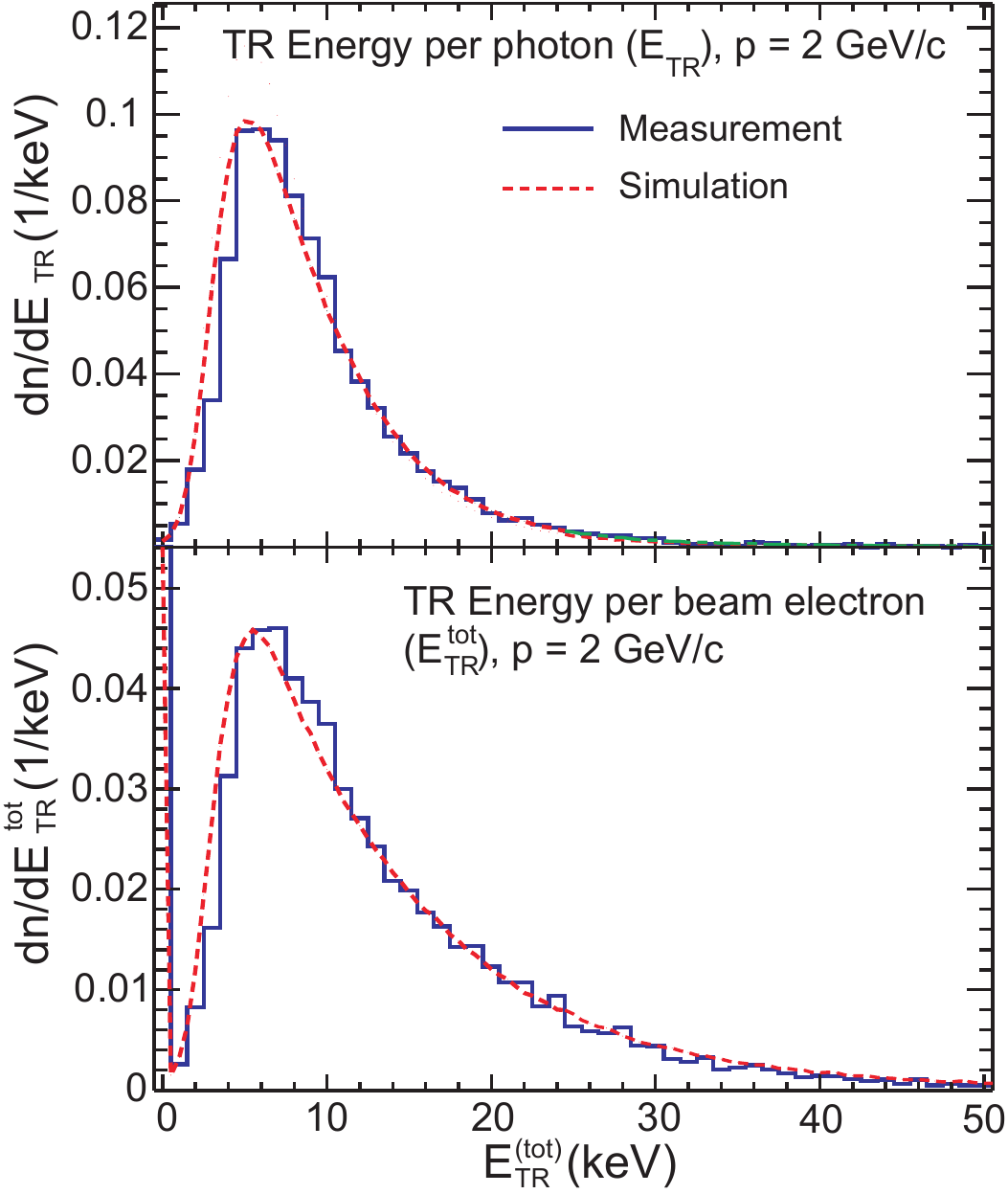}
\caption{Measurement and simulation of the transition radiation
  spectrum from 2 GeV/c electrons with the ALICE TRD
  radiator sandwich. The top panel shows the energy distribution of
  TR photons, the bottom panel the response per incident
  electron \cite{alice-tr}.}
\label{alice:tr} 
\end{minipage} \end{tabular}
\end{figure} 

Measurements of the specific energy loss of electrons and pions of 
2 GeV/c momentum in the Xe/CO$_2$ gas mixture are shown in
Fig.~\ref{alice:dedx} along with simulations. A correct understanding of
the particle separation capability relies on a precise
understanding of the details in the specific energy deposit of
electrons and pions. Fig.~\ref{alice:dedx} demonstrates the level of
agreement that has been achieved in the simulations. In that context
it has been shown that a correct description of the escape
probability of energetic $\delta$-electrons is needed to describe the
tails of the energy distributions as well as their
momentum dependence \cite{alice-dedx}.

The radiator is a composite structure using different inhomogeneous
materials. 
The front and back sides consist of 8 mm Rohacell foam
covered with 0.1 mm carbon fiber laminate and 25 $\mu$m aluminized
mylar foil. It is filled with irregular polypropylene fiber mats (average
fiber diameter 20 $\mu$m). 
The transition radiation production of this
structure has been evaluated in prototype tests employing a method
to separate TR from the parent track via electron deflection
in a magnetic field, as described in \cite{fab}. 
For electrons the production of transition radiation
sets in at $p\approx$0.5 GeV/c and levels off at
about 2 GeV/c, where on average 1.45 transition radiation
photons are produced of which 1.25 are detected per incident
electron\cite{alice-tr}. The measured transition radiation energy
spectra along with simulations are shown for 2 GeV/c electrons in 
Fig.~\ref{alice:tr}.

\subsubsection{Electron identification}
A parameterization of the measured amplitude spectra as a
function of drift time and momentum both for electrons and pions
provides the necessary likelihood distributions~\cite{alice-neural}
that allow to evaluate the electron identification performance.

\begin{figure}[hbt]
%\centering\mbox{\epsfig{file=./figs/eff-p-3meth.eps,width=0.6\textwidth}}
\centering\includegraphics[width=.6\textwidth]{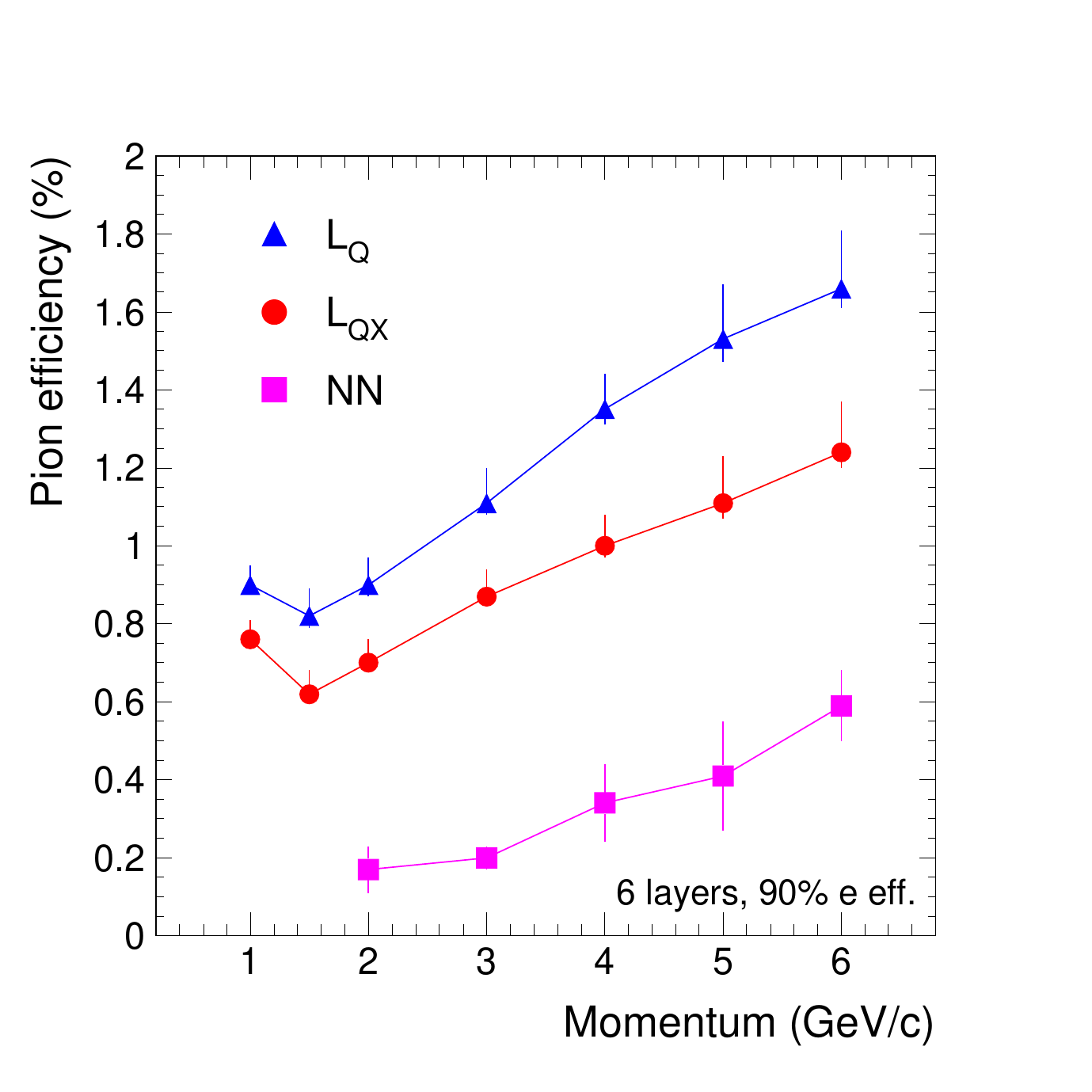}
\caption{Pion rejection as a function of momentum for three methods:
i) likelihood on integral charge (L$_Q$), ii) bidimensional likelihood 
on integral charge and largest cluster position (L$_{QX}$), 
iii) neural networks (NN). Figure taken from \cite{alice-ppr}.} 
\label{f:p_3met} 
\end{figure}

The dependence of the pion efficiency on momentum (for 90\% electron efficiency)
obtained with the ALICE TRD prototype measurements \cite{alice-ppr}
is shown in Fig.~\ref{f:p_3met} for three methods: 
i) likelihood on integral charge (L$_Q$), ii) bidimensional likelihood 
on integral charge and largest cluster position (L$_{QX}$), 
iii) neural networks (NN).
As expected, the higher performance of the more complex signal readout 
and processing is clearly demonstrated \cite{alice-neural}. 

\subsubsection{Tracking}
\begin{figure}[hbt]
\begin{tabular}{lr} \begin{minipage}{.65\textwidth}
\centering\includegraphics[width=.9\textwidth]{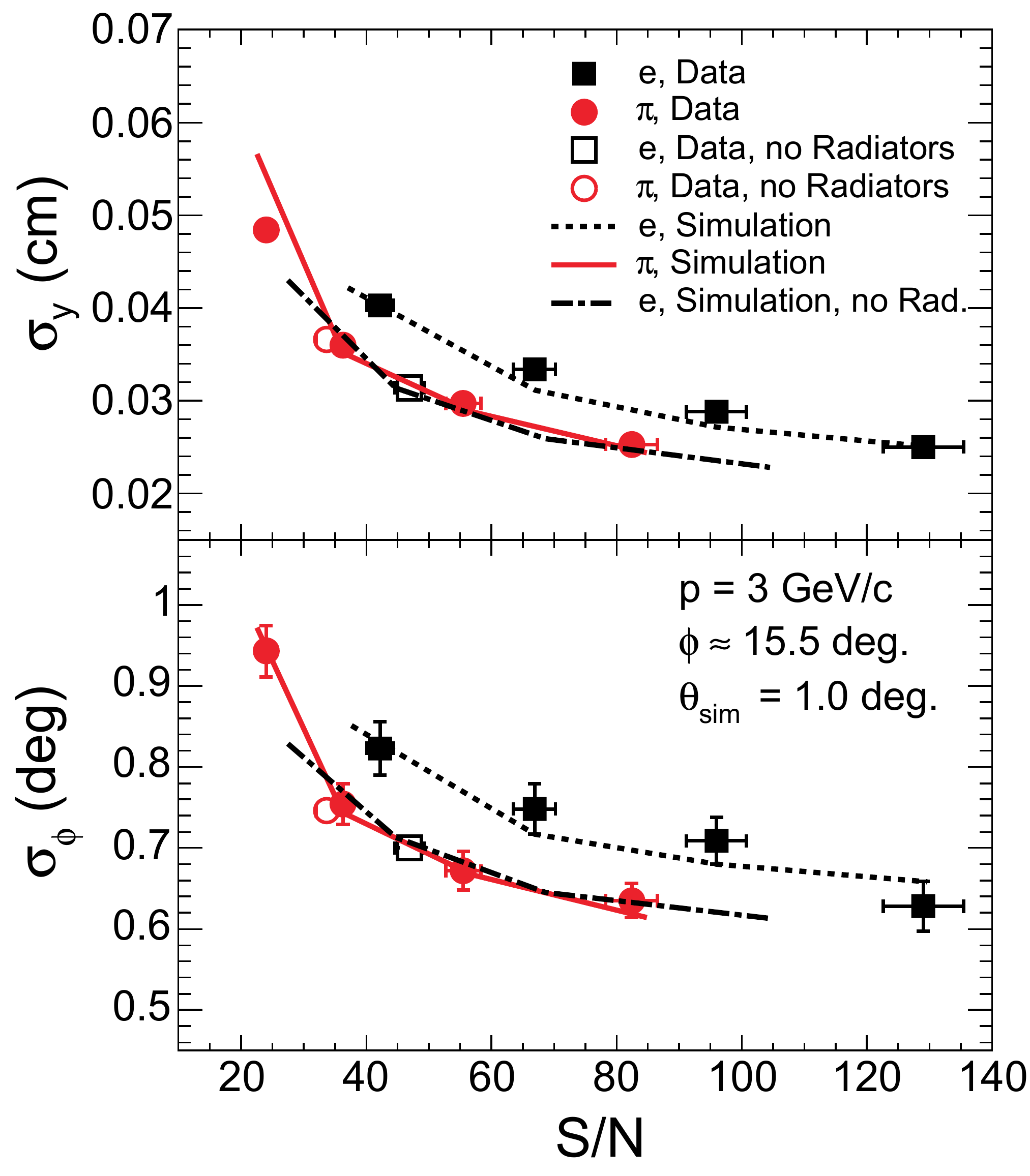}
\end{minipage} &\begin{minipage}{.32\textwidth}
\caption{Measurements and simulations of the position (top) and
  angular resolution of the ALICE TRD as a function of the signal-to-noise ratio.
  Open (full) symbols refer to electrons and pions measured with (without)
  radiator in front of the drift chamber along with the respective
  simulations \cite{alice-posres}.}\label{alice:posres} 
\end{minipage} \end{tabular}
\end{figure}

The chief tracking device in ALICE is the TPC. The tracking performance
of the ALICE TRD itself is a requirement regarding a reasonably sharp trigger
threshold for individual particles in the range of up to $p_{\,\rm t}
\approx 10$ GeV/c. At nominal magnetic field (B=0.5~T)
this entails a position resolution for each time bin of $\sigma_y
\,\lesssim\, 400\,\mu\rm m$ and a resulting angular resolution per
layer of $\sigma_\varphi \,\lesssim\, 1^\circ$. The achievable respective
resolutions have been measured and are a function of the
signal-to-noise ratio~\cite{alice-posres}. They are shown in
Fig.~\ref{alice:posres}. At a signal-to-noise ratio of about
40 the detector meets the requirements.

Using the above position and angular resolutions the stand-alone
tracking resolution of the TRD was estimated in simulations for different 
momenta as a function of multiplicity density. For momenta below 2 GeV/c 
the stand-alone momentum resolution of the TRD is around
$\delta p_{\,\rm t}/p_{\,\rm t} \approx 2.5-3\% $ with little
dependence on the multiplicity.
Through the inclusion of the TRD into the tracking in the central barrel an 
overall momentum resolution around 3\% can be obtained up to momenta of 
about 90 GeV/c.

\subsection{TRDs for fixed-target accelerator experiments}
A great variety of TRDs were employed for fixed-target high-energy experiments.
We discuss here, briefly, the TRD of the HERMES experiment at HERA \cite{her}
and that of the proposed CBM experiment at the future FAIR facility~\cite{cbm}.

\subsubsection{HERMES}\label{sect:hermes}

The TRD of the HERMES experiment \cite{her} employed random fiber radiators 
of 6.35~cm thickness (corresponding on average to 267 dielectric layers) and 
proportional wire chambers of 2.54~cm thickness, filled with Xe-CH$_4$. 
The TRD consisted of two arms, each with 6 radiator-detector layers
flushed with CO$_2$ in between. 
As a consequence of a rather thick radiator, the pion rejection factor 
achieved with a truncated mean method was 130 for a momentum of 5 GeV/c
and 150 averaged over all measured momenta, for an electron efficiency of 90\%.
Using a likelihood method, the pion rejection factor averaged over all
measured momenta
was determines to be 1460$\pm$150, decreasing to 489$\pm$25 for an
electron efficiency of 95\%.

\subsubsection{CBM}\label{sect:cbm}

The TRD of the CBM (Compressed Baryonic Matter) experiment \cite{cbm} 
at the planned FAIR \cite{aa:fair} accelerator facility at GSI
is aimed to provide electron identification and charged particle tracking. 
The required pion suppression is a factor of about 100 and the position 
resolution has to be of the order of 200-300 $\mu$m. In order to fulfill 
these tasks, in the context of the high rates and high particle multiplicities 
in CBM, a careful optimization of the detector is required.

Currently, the whole detector is envisaged to be subdivided into three
stations, positioned at distances of 4, 6 and 8 m from the target,
each one of them composed of at least three layers.
Because of the high rate environment expected in the CBM experiment
(interaction rates of up to 10 MHz), a fast readout detector 
has to be used. 
To ensure the speed and also to minimize possible space charge effects
expected at high rates, it is clear that the detector has to have
a thickness of less than 1~cm. Two solutions exist for such a
detector: a multiwire proportional chamber (MWPC) with pad readout
or straw tubes. 
While both had been investigated at the earlier stage of the detector design,
the MWPC solution is currently favored. A novel concept of
a ``double-sided'' MWPC had been tested in prototypes \cite{cbm-double}
and is a strong candidate for the inner part of the detector. This
detector design provides twice the thickness of the gas volume, while
keeping the charge collection time to that of a single MWPC.
For the radiator both possibilities, regular and irregular, are 
under consideration. The final choice of the radiator type for the CBM TRD 
will be established after the completion of prototypes tests.
Measurements with prototypes, both in beam \cite{cbm-rate}
and with X-ray sources \cite{cbm-rate2} demonstrate that the detector 
can handle the design rates.

The main characteristics of the TRD are: 
i) cell sizes: 1-10 cm$^2$ (depending on the polar angle, tuned for 
the occupancy to remain below 10\%);
ii) material budget: $X/X_0\simeq$15-20\%;
iii) rates: up to 100 kHz/cm$^2$;
iv) doses (charged particles): up to 16 krad/year, corresponding to
26-40 mC/cm/year charge on the anode wires.
For a classical MWPC-type TRD with the envisaged 9-12 layers, the
total area of detectors is in the range 485-646~m$^2$. The total number 
of electronic channels is projected to between 562 and 749 thousand.

\subsection{TRDs for astro-particle physics}

A recent review of TRDs for astro-particle instruments is given in \cite{tkirn}.
In general, both balloon and space experiments lead to compact design 
requirements.
For short-term balloon flights, like the WIZARD/TS93~\cite{wizard} and 
HEAT~\cite{heat} experiments, the main challenge is the rather strong variation 
of temperature and pressure during the flight, which require
significant corrections of 
the measured detector signals.
The requirements imposed by the long-term operation of a TRD in space as 
envisaged for the AMS experiment~\cite{ams}, lead to challenging aspects of 
its operation without maintenance. The mechanical requirements arising from 
vibrations during the launch demand special design and laboratory 
qualifications~\cite{danilo,tkirn}.

The TRD of the WIZARD/TS93 experiment~\cite{wizard} weighs about 240~kg and 
covers an active area of 76$\times$80~cm$^2$. Ten layers of carbon fiber 
radiators of 5~cm thickness and 1.6~cm-thick proportional wire chambers filled 
with Xe-CH$_4$ give a total of 2560 electronics readout channels. 
A pion contamination at the sub-percent level has been achieved in testbeam 
measurements.

The TRD of the HEAT experiment~\cite{heat} is composed of six layers of 
polyethylene fiber radiators (12.7~cm thickness) and 2~cm-thick proportional 
wire chambers operated with Xe-CH$_4$. Proton rejection factors around 100 
were achieved for 90\% electron efficiency for 10 GeV/c momentum.

The TRD designed for the PAMELA experiment~\cite{pam} is composed of a total 
of 1024 straw-tube detectors of 28~cm length and 4~mm diameter, 
filled with Xe-CO$_2$ mixture and arranged in 9 layers interleaved with radiators of carbon 
fibers. Pion rejection factors around 20 for 90\% electron efficiency
were measured in testbeams at momenta of few GeV/c.

\begin{figure}[htb]
\centering\includegraphics[width=.7\textwidth]{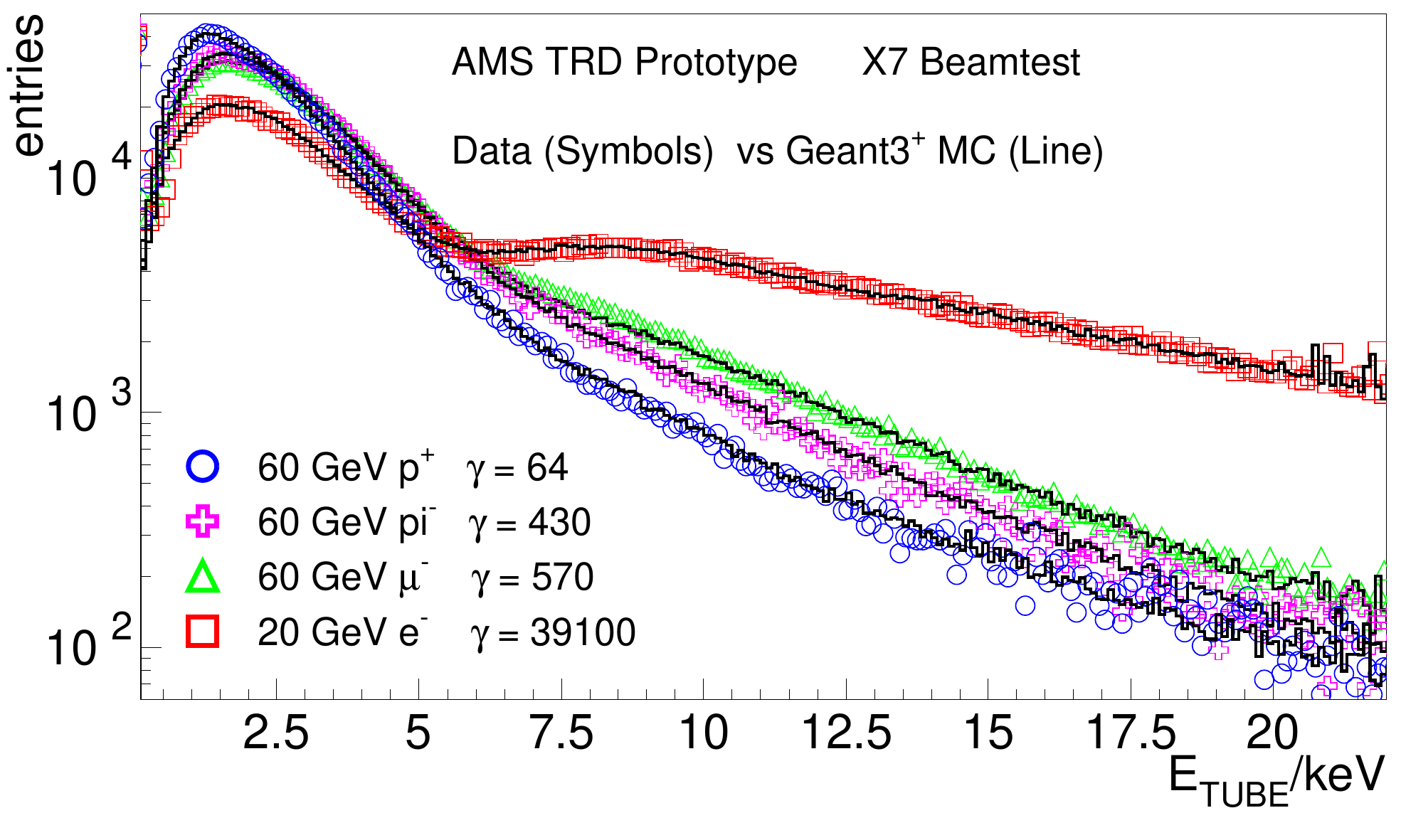}
\caption{Energy deposition in a single straw of the AMS TRD for protons, 
pions, muons and electrons obtained in beamtest measurements.
The lines are Geant3 simulations (from~\cite{ams}).}
\label{f:ams1}
\end{figure}

The TRD of the AMS experiment~\cite{ams}, which was recently installed
on the International Space Station (ISS), has an envisaged operational duration
of about three years. The TRD will contribute to the AMS required proton 
rejection factor of about 10$^6$, necessary for the study of positron spectra 
planned with AMS.
The detection elements are 5248 straw tubes of 6~mm diameter, arranged in 
modules of 16 straws each, with a length of up to 2~m.
The straws, with 30~$\mu$m gold-plated tungsten anode wires, are operated at 
1350~V, corresponding to a gas gain of 3000.

The radiator is a 2~cm thick polypropylene fleece. Special cleaning of the 
radiator material is required to meet the outgassing limits imposed by NASA.
Special tightness requirement for the straw tubes~\cite{tkirn2} are imposed by 
the limited supply of detector gas (the AMS TRD has a gas volume of 230 liters).

The spectra of energy deposition in a single straw of the AMS TRD for protons, 
pions, muons and electrons obtained in beamtest measurements are shown in
Fig.~\ref{f:ams1}. A very good description of the measurements has been achieved
with modified Geant3 simulations~\cite{ams}.

\begin{figure}[htb]
\centering\includegraphics[width=.7\textwidth]{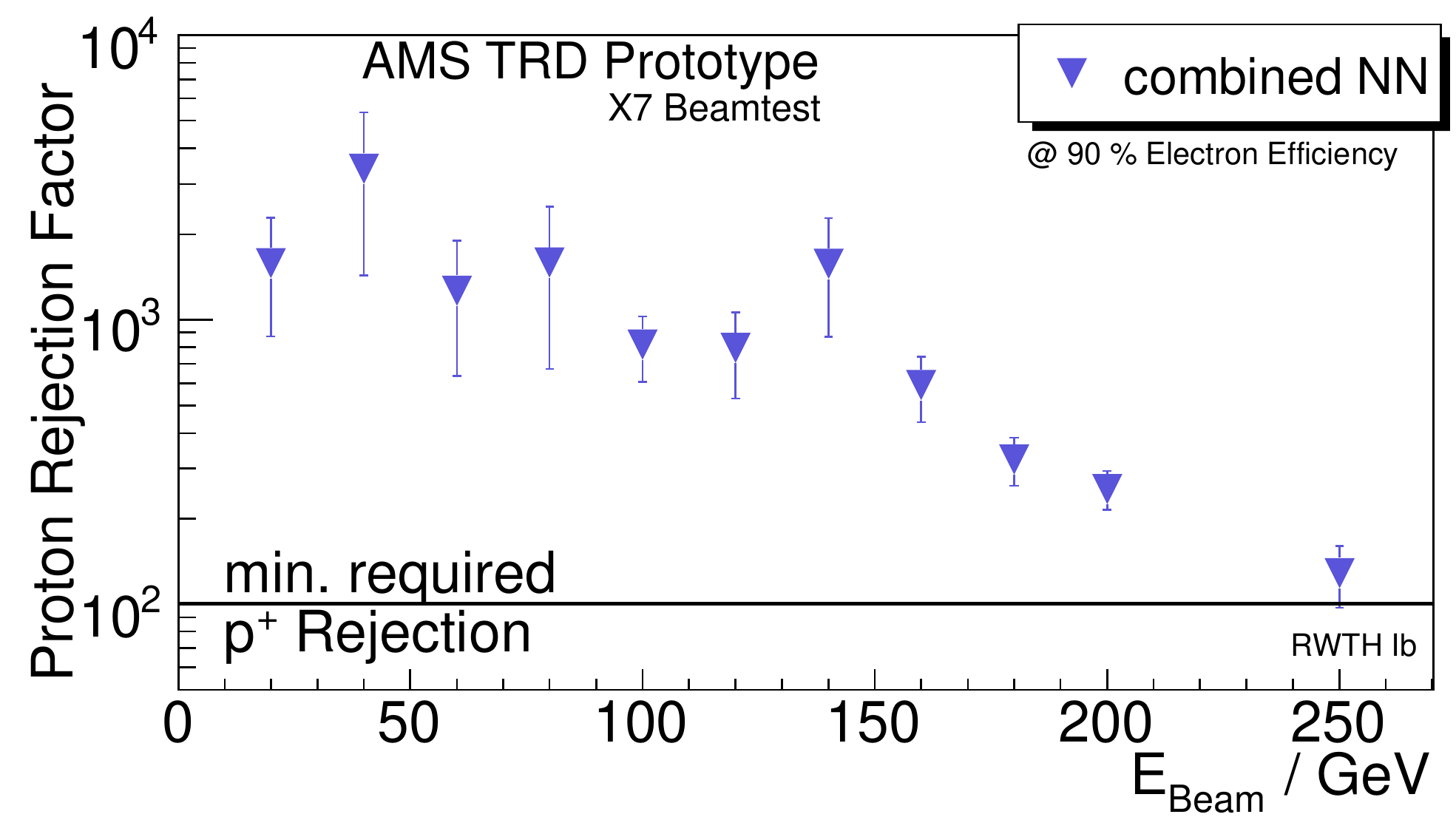}
\caption{The proton rejection factor of the AMS-02 TRD as a function of 
beam energy (from~\cite{ams}).}
\label{f:ams2}
\end{figure}

The excellent proton rejection performance achieved in testbeams with a full 
20 layer prototype for the AMS TRD is shown in Fig.~\ref{f:ams2}. 
A neural network method has been used~\cite{ams}, delivering a proton rejection 
factor (at 90\% electron efficiency) between 1000 and 100 for momenta between 
15 and 250~GeV/c.

\section{Summary and conclusions}\label{sect:sum}

The TRD technique offers a unique opportunity for electron separation with
respect to hadrons in a wide momentum range from 1 to 100 GeV/c.
The separation between pions, kaons and protons (or heavier hadrons) is possible 
in well defined windows of momenta.
We have presented a survey of the Transition Radiation Detectors employed
in accelerator and space experiments, with special emphasis on the two large 
detectors presently operated in the LHC experiments, the ATLAS TRT and the 
ALICE TRD. 
Building on a long series of dedicated measurements and on various implementation
of TRDs in complex experimental particle physics setups, these two particular
TRD systems are challenging in their scale and required performance, both for 
tracking and electron identification. The ALICE TRD provides in
addition fast triggering 
capability.
They also illustrate two complementary approaches, dictated by their respective
requirements: the ATLAS TRT being a very fast detector, with moderate 
granularity, perfectly suited for operation in high-rate pp collisions, 
while the ALICE TRD is a slower detector with very good granularity, 
optimized for Pb+Pb collisions.
With data taking at the LHC now in full swing, the evaluation of the performance
of these two systems, which is already well underway \cite{perf_lhc}, will 
serve as a solid basis for the design of TRDs for future high-energy 
(astro-)particle and nuclear physics experiments.

\end{document}